%% file: main.tex
\documentclass[10pt,conference]{IEEEtran} 
\IEEEoverridecommandlockouts
\usepackage{cite}
\usepackage{amsmath,amssymb,amsfonts}
\usepackage{algorithmic}
\usepackage{graphicx}
\usepackage{textcomp}
\usepackage{xcolor}
\def\BibTeX{{\rm B\kern-.05em{\sc i\kern-.025em b}\kern-.08em
    T\kern-.1667em\lower.7ex\hbox{E}\kern-.125emX}}
    
\usepackage{fancyhdr}
\input{tool}
\usepackage[hyphens]{url}  
\usepackage{mathrsfs}
\usepackage{tcolorbox}
\newcommand{\boxmargin}{0.1mm}
\tcbset{colback=gray!8,
        colframe=black,
        width=8.7cm,
        arc=2mm, auto outer arc,
        boxrule = 1.0pt,
        left = \boxmargin, right = \boxmargin, top = \boxmargin, bottom = \boxmargin,
        leftright skip=0.5mm
}
\begin{document}

\title{CoCoSoDa: Effective Contrastive Learning for Code Search}

\newcommand\corrauthorfootnote[1]{%
  \begingroup
  \renewcommand\thefootnote{}\footnote{\textsuperscript{\S}#1}%
  \addtocounter{footnote}{-1}%
  \endgroup
}

\newcommand\notedauthorfootnote[1]{%
  \begingroup
  \renewcommand\thefootnote{}\footnote{\textsuperscript{\dag}#1}%
  \addtocounter{footnote}{-1}%
  \endgroup
}
\author{
Ensheng Shi\textsuperscript{a,\dag}
Yanlin Wang\textsuperscript{b,\S}
Wenchao Gu\textsuperscript{c,\dag}
Lun Du\textsuperscript{d}\\
Hongyu Zhang\textsuperscript{e}
Shi Han\textsuperscript{d}
Dongmei Zhang\textsuperscript{d}
Hongbin Sun\textsuperscript{a,\S}
\\
\textsuperscript{a}Xi'an Jiaotong University \quad
\textsuperscript{b}School of Software Engineering, Sun Yat-sen University \\
\textsuperscript{c}The Chinese University of Hong Kong \quad 
\textsuperscript{d}Microsoft Research  \quad
\textsuperscript{e}Chongqing University \quad

\\
{ s1530129650@stu.xjtu.edu.cn, wangylin36@mail.sysu.edu.cn}\\
{  wcgu@cse.cuhk.edu.hk, \{lun.du, shihan, dongmeiz\}@microsoft.com}\\
{ hyzhang@cqu.edu.cn, hsun@mail.xjtu.edu.cn } \\
}
\maketitle

\begin{abstract}
Code search aims to retrieve semantically relevant code snippets for a given natural language query. Recently, many approaches employing contrastive learning have shown promising results on code representation learning and greatly improved the performance of code search. 
However, there is still a lot of room for improvement in using contrastive learning for code search. In this paper, we propose CoCoSoDa to effectively utilize contrastive learning for code search via two key factors in contrastive learning: data augmentation and negative samples. Specifically, soft data augmentation is to dynamically masking or replacing some tokens with their types for input sequences to generate positive samples. Momentum mechanism is used to generate large and consistent representations of negative samples in a mini-batch through maintaining a queue and a momentum encoder. In addition, multimodal contrastive learning is used to pull together representations of code-query pairs and push apart the unpaired code snippets and queries. We conduct extensive experiments to evaluate the effectiveness of our approach on a large-scale dataset with six  programming languages. Experimental results show that: (1) CoCoSoDa outperforms \revise{18} baselines and especially exceeds CodeBERT, GraphCodeBERT, and UniXcoder by 13.3\%, 10.5\%, and 5.9\% on average MRR scores, respectively. (2) The ablation studies show the effectiveness of each component of our approach. (3) We  adapt our techniques to several different pre-trained models such as RoBERTa, CodeBERT, and GraphCodeBERT and observe a significant boost in their performance in code search. (4) Our model performs robustly under different hyper-parameters. Furthermore, we perform qualitative and quantitative analyses to explore reasons behind the good performance of our model. 
\end{abstract}


\section{Introduction}
Code search\corrauthorfootnote{Yanlin Wang and Hongbin Sun are the corresponding authors.} plays an important role in software development and maintenance~\cite{SingerLVA97,NieJRSL16}. To implement a certain functionality\notedauthorfootnote{Work done during the author’s employment at Microsoft Research Asia}, developers often search and reuse previously-written  code from open source repositories such as GitHub or from a large local codebase~\cite{McMillanGPXF11,Reiss09,ZhangNKZ18}. 
The key challenge in this task is to find semantically relevant code snippets \emph{written in programming languages} 
based on input queries \emph{written in natural languages}~\cite{AllamanisBDS18}.

Early studies \cite{LinsteadBNRLB09,LuSWLD15,LvZLWZZ15,McMillanGPXF11} on code search mainly leverage on the lexical information of the code snippets and use information retrieval (IR) methods. Later on, deep learning-based approaches \cite{GuZ018,DuSWSHZ21,CambroneroLKS019,LingWWPMXLWJ21,LiQYSC20,ZhuSLXZ20,ShuaiX0Y0L20,YeXZHWZ20,HaldarWXH20,GuCM21,WanSSXZ0Y19,LingLZX20} that employ neural networks to embed code and queries into a shared embedding space and measure their semantic similarity through vector distances are explored. Recently, large pre-trained code models~\cite{GuoRLFT0ZDSFTDC21,FengGTDFGS0LJZ20,GuoLDW0022,0034WJH21,AhmadCRC21}, which are pre-trained on large multi-programming-language data, improve the understanding of code semantics and achieve better code search performance. Some studies~\cite{0001JZA0S21,BuiYJ21} apply contrastive learning to unsupervised code representation learning and \revise{\sigircon~\cite{BuiYJ21}} also achieves good performance in code search. In detail, they first use semantic-preserving transformations~\cite{BuiYJ21} to generate similar code snippets as positive samples. These transformations are including \texttt{Statements Permutation}~\cite{BuiYJ21} (swapping two statements that have no data dependency on each other in a basic block  ), \texttt{Loop  Exchange}~\cite{BuiYJ21} (replacing \texttt{for} statements with \texttt{while} statements or vice versa.), etc. Second, they treat other code snippets in the same mini-batch as negative samples and then optimize the model to pull together representations of positive samples and push apart representations of negative samples. These approaches have shown promising results in code search. For example, \sigircon~\cite{BuiYJ21} outperforms all approaches and  achieve a MRR score of 0.727 on CodeSearchNet~\cite{Husain2019} Java dataset. However, there is still a lot of room for improvement in leveraging contrastive learning for code search, such as using other effective data augmentation to generate positive samples or enriching negative samples.

\begin{figure*}[t]
    \centering
    \includegraphics[width=0.95\linewidth]{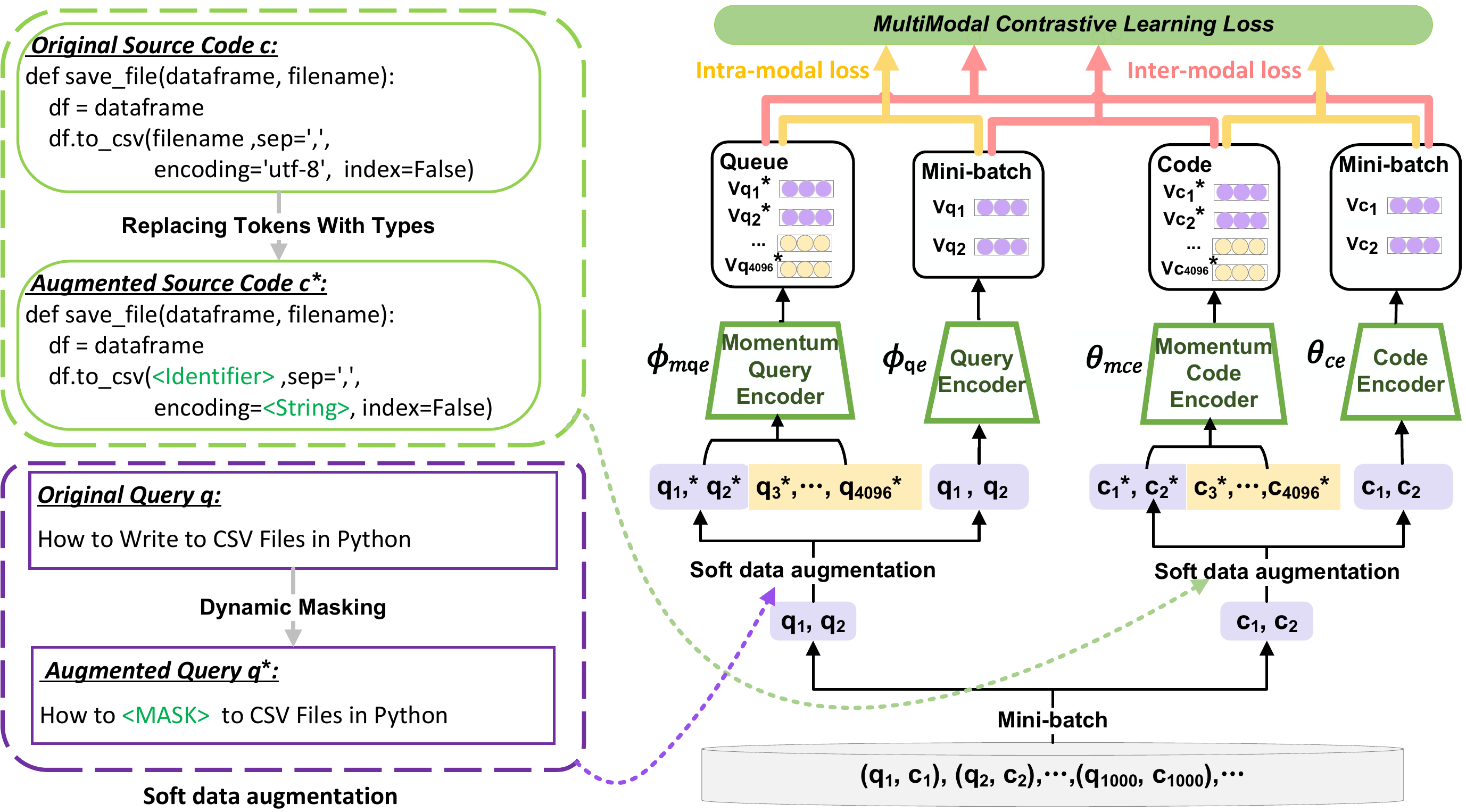}
      \caption{The framework of \Our{}.} 
    \label{fig:framework}
    \vspace{-3pt}
\end{figure*}

In this paper, we propose \textbf{\Our{}} (stands for \underline{Co}de search with multimodal momentum \underline{Co}ntrastive learning and \underline{So}ft \underline{D}ata \underline{a}ugmentation) {to effectively utilize contrastive learning for code search} via two key factors in contrastive learning: data augmentation and negative samples. \textcircled{1} To learn a better overall semantic representation of the code snippet and query instead of focusing on token-level semantic representation learning according to the surrounding context, we propose four SoDa (stands for \underline{So}ft \underline{D}ata \underline{a}ugmentation) methods (\Sec{}~\ref{sec:soda}). They dynamically replace $r$ ($r$ is stable from 5\% to 20\%) of code tokens with their types or simply mask them to generate similar code snippets. \textcircled{2} To distinguish one sample 
from more negative samples at each iteration, we adopt the momentum mechanism~\cite{He0WXG20} to enlarge negative samples in a mini-batch. \textcircled{3} We also employ multimodal contrastive learning to minimize the distance between the representations of code-query pair and maximize the distance between the representations of the query (code snippet) and other many unpaired code snippets (queries). \revise{The overall framework of \Our{} is shown in~\Fig~\ref{fig:framework}. On the left is an example of soft data augmentation, and on the right is the main architecture of our model. More details are given in \Sec{}~\ref{our_method}}.

We evaluate the effectiveness of \Our{} on a large-scale dataset CodeSearchNet~\cite{Husain2019} with six programming language (\ruby{}, \javascript{}, \go, \python, \java, \php) and compare \Our{} with \revise{18} state-of-the-art (SOTA) approaches. We also conduct the ablation study to study the effectiveness of each component of \Our{}. Furthermore, we apply \Our{} to other three large-scale pre-trained models, including natural language pre-trained model \roberta~\cite{Liu2019Roberta}, code pre-trained models \codebert~\cite{FengGTDFGS0LJZ20} and \graphcodebert~\cite{GuoRLFT0ZDSFTDC21}. We also assign different hyperparameters to check their impact on code search. In addition, we discuss why \Our{} perform well through qualitative and quantative analyses.
Experimental results show that: \textit{(1)} \Our{} significantly outperforms existing SOTA approaches on code search task (\Sec{}~\ref{RQ1}). \textit{(2)} The multimodal momentum contrastive learning including intra-modal and inter-modal contrastive learning and  four SoDa methods play important roles individually and can improve the performance of the code search model (\Sec{}~\ref{RQ2:ablation-study}).
 \textit{(3)} \Our{} can be easily adapted to other pre-trained models and obviously boost their performance (\Sec{}~\ref{RQ_other_pre_train}).
\textit{(4)} \Our{} performs stably over a range of hyperparameters: learning rate is from $5e^{-6}$ to $7e^{-5}$, momentum coefficient $m$ is between 0.910 and 0.999, masked ratio $r$ is from 5\% to 20\%, and temperature hyperparameter $\tau$ varies from 0.03 to 0.07 (\Sec{}~\ref{RQ_exp_hyperparameter}). 

We summarize the contributions of this paper as follows:
\begin{itemize}
     \item We adapt Transform-based momentum contrastive learning algorithm to better leveraging contrastive learning techniques for code search task. It enables the model to learn effective code representations by learning better representation of one sample against more negative samples. We also propose a new approach incorporating multimodal momentum contrastive learning. It can pull together the representations of matched code-query pairs and push apart the representations of unmatched code-query pairs. 
    
    \item We propose four simple yet effective soft data augmentation methods that utilize dynamic masking and replacement for data augmentation. More importantly, \soda can be easily applied to all programming languages.

    \item We conduct extensive experiments to evaluate the superiority of our approach on a large-scale multi-programming-language dataset. The results show that our approach significantly outperforms baselines, our framework can be easily adapted to other pre-trained models and significantly boost their performance, and \Our{} performs stably over a range of hyperparameters.

\end{itemize}

\section{Related Work}
\label{related_work}
\subsection{Code Search}

\revise{Learning the representation of code is an emerging topic and has been found to be useful in many software engineering tasks, such as code summarization~\cite{IyerKCZ16,zhangretrieval20,LeClairJM19,shi2021cast,shia2022evaluation}, code search~\cite{ZhuSLXZ20,HaldarWXH20,GuZ018,du2021single,gu2022accelerating}, code completion~\cite{abs-2004-13651,ProkschLM15,RaychevVY14,BruchMM09,wang2021code}, commit message generation~\cite{tao2022large,VasquezCAP15,Xu00GT019,tao2021evaluation,shi-etal-2022-race}. 
Among them,} code search plays an important role in software development and maintenance~\cite{SingerLVA97,NieJRSL16}.
Traditional approaches~\cite{LinsteadBNRLB09,LuSWLD15,LvZLWZZ15,McMillanGPXF11} based on retrieval information techniques mainly focus on the lexical information of the source code and apply keywords matching methods to search the relevant 
code snippets for the given query.
In recent years, deep learning-based approaches
leverage the neural network to learn the semantic representations of the source code and natural language to improve the understanding of code snippets and queries.
Gu et al.~\cite{GuZ018} is the first to use the deep neural network to embed the code and query into a shared vector space and measure the similarity of them using vector distance. Subsequently, various types of model structures are applied to code search, including sequential
models~\cite{GuCM21,HaldarWXH20,ShuaiX0Y0L20,WanSSXZ0Y19,YeXZHWZ20}, convolutional neural network~\cite{LiQYSC20,LingLZX20,ZhuSLXZ20}, tree neural network~\cite{WanSSXZ0Y19}, graph models~\cite{LingWWPMXLWJ21,WanSSXZ0Y19}, and transformers~\cite{DuSWSHZ21,ZhuSLXZ20}. Recently, large-scale code pre-trained models~\cite{GuoRLFT0ZDSFTDC21,FengGTDFGS0LJZ20,GuoLDW0022,0034WJH21,AhmadCRC21,NiuL0GH022}, which are pre-trained on a massive source code dataset, improve the understanding of code semantics and achieve significant improvements in code search task. For example, \codebert is pre-trained with~\revise{masked language modelling (MLM)}, which is to predict masked tokens, and~\revise{replaced token detection (RTD)}, which uses a discriminator to identify replaced tokens. \graphcodebert takes source code, paired summarization and the corresponding data flow as the input and is pre-trained with MLM, data flow edge prediction, and node alignment tasks. \plbart~\cite{AhmadCRC21} is a  sequence-to-sequence code pre-trained models and is pre-trained with denoising autoencoding, which destroys a span of tokens and then recovers them. Our approach can be easily applied to these pre-trained models and boost their performance.

\subsection{ Code Representation Learning with Contrastive Learning}
Contrastive learning approaches~\cite{HadsellCL06}, which pull close the similar representations and push apart different representations, have been successfully used in self-supervised representation learning on images~\cite{He0WXG20,ChenK0H20} and natural language texts~\cite{GaoYC21,Fang2020,GiorgiNWB20}. 
To generate individual augmentations, images usually use spatial~\cite{GidarisSK18,Devries2017} and appearance transformation~\cite{Howard13,IoffeS15}, and natural language texts mostly use back-translation approach~\cite{Fang2020} and spans technique~\cite{GiorgiNWB20}. Then, a model is pre-trained to identify whether the augmented samples are from the same original sample. 

Recently, some studies~\cite{0001JZA0S21,BuiYJ21,wang2021syncobert,Ding2021,GuoLDW0022} try to use contrastive learning approaches on unsupervised code representation learning. 
For example, Jain et al. ~\cite{0001JZA0S21} and Bui et al. ~\cite{BuiYJ21} mainly use semantic-preserving program transformations to generate the functionally equivalent code snippets and pre-train the model to recognize semantically equivalent and non-equivalent code snippets through contrastive learning techniques. These transformations including variable renaming (rename a variable with a random token), unused statement (insert an dead code snippet such as an unused declaration statement), permute statement (swap two statements which have no data dependency on each other), etc.
Unlike the above-mentioned pre-trained technique, 
our model is based on multimodal contrastive learning with momentum encoders, which allow the model to learn the good representation based on samples in the current mini-batch and previous mini-batches. Furthermore, previous data augmentations require  preserving the semantics of source code, whereas we use a simple yet effective dynamic masking technique that allows more flexible soft data augmentation. 

\section{Proposed Approach}
\label{our_method}

\revise{
In this section, we illustrate our model \Our{} for code search. The overall framework of \Our{} is shown in \Fig~\ref{fig:framework}. It is comprised of the following components:
\begin{itemize}
\item \textbf{Pre-trained code/query encoder} captures the semantic information of a code snippet or a natural language query and maps it into a high-dimensional embedding space. 

\item \textbf{Momentum code/query encoder} encodes the samples (code snippets or queries) of current and previous mini-batches to enrich the negative samples.

\item \textbf{Soft data augmentation} is to dynamically mask or replace some tokens in a sample (code/query) to generate a similar sample as a form of data augmentation. 

\item \textbf{Multimodal contrastive learning} is used as the optimization objective and consists of inter-modal and intra-modal contrastive learning loss. They are used to minimize the distance of the representations of similar samples and maximize the distance of different samples in the embedding space.

\end{itemize}
We first illustrate our model with a concrete example shown in \Fig~\ref{fig:framework} and then introduce each component in details.

\subsection{An Illustrative Example}
In this section, we introduce our model by an illustrative example shown in~\Fig~\ref{fig:framework}. On the left side is an example of soft data augmentation performing on the code snippet and query pair, and on the right side is the main architecture of our model. Specifically, \textit{first}, at each iteration, we randomly perform one of four SoDa methods including \Masking{} (\ShortMasking), \Replacement (\ShortReplacement), \TypeReplacement (\ShortTypeReplacement), and \TypeMasking (\ShortTypeMasking), to generate the positive samples. \ShortMasking and \ShortReplacement randomly sample 15\% of tokens of a code snippet and replace each token with a \texttt{[MASK]} token or the type of the token. \ShortTypeReplacement and \ShortTypeMasking sample all tokens of the specified type (such as operator, identifier) from a code snippet, and 15\% of them are randomly replaced with the specified type  or \texttt{[MASK]} token. For query, only \ShortMasking is performed because other three \soda methods require the type information of source code. 

\textit{Second}, we adopt the framework of momentum contrastive learning (MoCo)~\cite{He0WXG20}, and apply the multi-layer Transformer encoder~\cite{VaswaniSPUJGKP17} as the backbone of the (momentum) encode because Transformer can effectively model and represent source code~\cite{FengGTDFGS0LJZ20,GuoRLFT0ZDSFTDC21}. We further extend MoCo for multimodal representation learning by doubling the encoder and momentum encoder. Next, we use a large-scale code pre-trained model \unixcoder{}~\cite{GuoLDW0022} to initialize the code/query encoder and momentum code/query encoder (\Fig~\ref{fig:framework}), and then feed the original and augmented samples (codes or queries) to the encoders and momentum encoders, respectively, to obtain representations of samples. The momentum encoder can generate large and consistent representations of negative samples compared to the common encoder updated by back-propagation and the fixed encoder~\cite{He0WXG20}. Therefore, at each iteration, \Our{} can be trained to distinguish one sample from more other negative samples, so our approach can learn better representations of code snippets and queries.

\textit{Third}, the multimodal contrastive learning consists of an intra-modal loss and an inter-modal loss is used to pull close  similar representations and push apart dissimilar representations. Specifically, the intra-modal loss is used to learn an uniform distribution of representations of unimodal data by pulling in similar samples (codes or queries) and pushing away different samples. The inter-modal is use to learn the alignment of the multimodal data by pull close  representations of the paired code snippet and query and push apart representations of the unpaired code snippet and query. 

\textit{Finally}, the well-trained model is used for code search. In detail,  the code and query encoders map code snippets of codebase and the given query into a high-dimensional space, measure the similarity between them with cosine similarity, and return the most relevant code snippet based on the similarity.
}

\subsection{Pre-trained Encoder and Momentum Encoder}

In this section, we introduce the base architecture, input samples, output representation and update mechanism of encoder and momentum encoder. The encoders and momentum encoders are all built on the multi-layer bidirectional Transformer Encoder~\cite{VaswaniSPUJGKP17}.
As the pre-trained models such as \unixcoder~\cite{GuoLDW0022} have achieved substantial improvement in code search, we initialize code and query encoder with parameters of \unixcoder. 
Following previous study~\cite{GuoLDW0022}, we average all the hidden states of the last layer as the whole sequence-level representation of query/code. 

In the MoCo~\cite{He0WXG20} framework, there is a momentum encoder encoding the samples of the current and previous mini-batches. Specifically, the momentum encoder maintains a queue by enqueuing the samples in the current mini-batch and dequeuing the samples in the oldest mini-batch. Here, we also use \unixcoder to initialize parameters of momentum code and query encoder. The difference of the update mechanism between the encoder and momentum encoder is that the encoder is updated by the back-propagation algorithm while the momentum encoder is updated by linear interpolation of the encoder and the momentum encoder. 
Thus, compared with the memory bank approach~\cite{WuXYL18}, which fixes and saves the representations of all samples of the training dataset in advance, the momentum encoder can generate consistent representations and has been demonstrated to be effective~\cite{He0WXG20}. 
For the end-to-end approach~\cite{GuCM21,GuoRLFT0ZDSFTDC21}, 
it has one encoder and takes other samples in the current mini-batch as negative samples. Thus, it requires a large mini-batch size in order to expand the number of negative samples~\cite{He0WXG20}.
For example, under the same computational resource such as  A100-PCIE-80GB~\cite{A100}, up to 199 negative samples can be used in a mini-batch for end-to-end approaches, but our approach can use over 4,000 negative samples. Therefore, to support more negative samples, end-to-end approaches require larger memory computational resources than our approach.

We denote the parameters of the code encoder as $\theta_{ce}$ and the momentum code encoder as $\theta_{mce}$, with parameters being the weights of \unixcoder. Therefore, $\theta_{mce}$ is updated
by:
\begin{equation}
\small
    \theta_{mce} = m\theta_{mce} + (1-m)\theta_{ce}
\end{equation}
\noindent where $m \in [0, 1) $ is a momentum coefficient. Similarly, we denote the parameters of the query encoder and moment query encoder as $\phi_{qe}$ and $\phi_{mqe}$. Then $\phi_{mqe}$ is updated
by:
\begin{equation}
\small
\phi_{mqe} = m\phi_{mqe} + (1-m)\phi_{qe}
\end{equation}
Both $\theta_{ce}$ and $\phi_{qe}$ are learnable parameters and updated by the back-propagation algorithm.

\subsection{Soft Data Augmentation}\label{sec:soda}

In this section, we introduce soft data augmentation (SoDa) methods, which are simple data augmentation approaches without external constraints for source code or queries. We first introduce how to obtain soft data augmentation and then introduce how to use the augmented data. 

The four SoDa methods are shown as follows.
\begin{itemize}
    \item Dynamic Masking (\ShortMasking): randomly sampling 15\% of tokens of a code snippet and replace each token with a [MASK] token.
    \item Dynamic Replacement (\ShortReplacement):  randomly sampling 15\% of tokens of a code snippet and replace each token with the type of the token.
    \item Dynamic Replacement of Specified Type (\ShortTypeReplacement): sampling all tokens of a specified type (such as operator, identifier) from a code snippet, and 15\% of them are randomly replaced with the specified type.
    \item Dynamic Masking of Specified Type (\ShortTypeMasking): sampling all tokens of a specified type (such as operator, identifier) from a code snippet, and 15\% of them are randomly masked.
\end{itemize}

Here, \textit{dynamic} means that in data processing, the masking or replacement operation is performed at each iteration rather than only performed once~\cite{Liu2019Roberta}. It is worth noting that we randomly perform one of four SoDa methods for code snippets at each iteration. In addition, only \ShortMasking is performed for queries because other three SoDa methods require the type information of source code.

We denote the SoDa module as $G_{soda}$ which performs data transformation operations for the given input sequence to obation the augmented data. Specifically, we first perform one of the four \soda methods for the code snippets $C = (c_1,...,c_{bs})$ and queries $Q= (q_1,...,q_{bs})$ in a mini-batch by:
\begin{equation}
\small
\begin{aligned}
    c_i^{\ast} = G_{soda} (c_i), \quad 
    q_i^{\ast} = G_{soda} (q_i) \quad (i=1,...,bs)
\end{aligned}
\end{equation}
where $c_i^{\ast}$ and $ q_i^{\ast}$ are the augmented samples of the code snippet $c_i$ and query $q_i $
, respectively. $bs$ is mini-batch size. Then code snippet $c_i$ and query $qi$ are fed into the code/query encoder and augmented samples $c_k^{\ast}$  and $ q_k^{\ast}$ ( $k=1,...,K$ and $K$ is the queue size) in the current and previous mini-batches are fed to the momentum code/query encoder by:
\begin{equation}
\small
\begin{aligned}
  \mathbf{v_{c_i}} &= f_{\theta_{ce}} (c_i), \quad \mathbf{v_{c_k^{\ast}}} = f_{\theta_{mce}} (c_k^{\ast}) \\
 \mathbf{ v_{q_i}} &= f_{\theta_{qe}} (q_i), \quad \mathbf{ v_{q_k^{\ast}}} = f_{\theta_{mqe}} (q_k^{\ast}) \\
\end{aligned}
\end{equation}

\noindent where, $\mathbf{v_{c_i}}, \mathbf{ v_{q_i}} ,\mathbf{v_{c_k^{\ast}}}$, and $\mathbf{ v_{q_k^{\ast}}}$ are the final overall representations of the code snippet $c_i$, query $q_i$, augmented code snippet $c_k^{\ast}$, and augmented query $q_k^{\ast}$, respectively. 

\subsection{Multimodal Contrastive Learning}
Multimodal contrastive learning consists of inter-modal and intra-modal loss function, and is used to optimize the parameters of the model. 
Specifically, given a query $q_i$, we denote the paired $c_i$ or $c_i^{\ast}$ as $c_i^+$ and unpaired $c_k^{\ast}$ as $c_k^-$ ($i=1,...,bs$ and  $k=1,...,K$). 
For the query $q_i$, with similarity measured by cosine similarity ($sim (\mathbf{x} ,\mathbf{y} ) = \frac{\mathbf{x \cdot y}}{\Vert\mathbf{x} \Vert \Vert \mathbf{y}\Vert }$), we define the inter-modal and intra-modal contrastive learning loss~\cite{oord2018representation,WanSSXZ0Y19} as:
\begin{equation}
\small
\begin{aligned}
   L^{inter}_{q_i} &= - \log \frac{ e^{ (sim (\mathbf{ v_{q_i}}, \mathbf{ v_{c_i}^+}) / \tau )} }{e^{(sim(\mathbf{ v_{q_i}},\mathbf{ v_{c_i}^+} )/\tau)}  + \sum\limits_{k=1}^{K} e^{ (sim (\mathbf{ v_{q_i}}\cdot \mathbf{ v_{c_k}^-}) / \tau)} } \\
L^{intra}_{q_i} &= - \log \frac{ e^{(sim (\mathbf{ v_{q_i}}, \mathbf{ v_{q_i}^+}) / \tau )} }{e^{ (sim(\mathbf{ v_{q_i}},\mathbf{ v_{q_i}^+} )/\tau)} + \sum\limits_{k=1}^{K} e^{ (sim (\mathbf{ v_{q_i}}\cdot \mathbf{ v_{q_k}^-}) / \tau)} }
\end{aligned}
\end{equation}
\noindent where $\tau$ is the temperature hyperparameter~\cite{WuXYL18,He0WXG20} and is set to 0.07 following previous works~\cite{He0WXG20,0001JZA0S21}. Intuitively, the optimization objective of inter-modal loss function is to maximize the semantic similarity of the query and its paired code snippet and minimize the semantic similarity of the query and its unpaired code snippets. The intra-modal loss function is to learn the better representations of queries, where similar queries have closed representations and different queries have distinguishing representations. In the same way, for a code snippet $c_i$, we define the corresponding multimodal contrastive learning loss as: 
\begin{equation}
\small
\begin{aligned}
    L^{inter}_{c_i} = - \log \frac{ e^{ (sim (\mathbf{ v_{c_i}}, \mathbf{ v_{q_i}^+}) / \tau ) }}{e^{ (sim(\mathbf{ v_{c_i}},\mathbf{ v_{q_i}^+} )/\tau)} + \sum\limits_{k=1}^{K} e^{ (sim (\mathbf{ v_{c_i}}\cdot \mathbf{ v_{q_k}^-}) / \tau) }} \\
L^{intra}_{c_i} = - \log \frac{ e^{ (sim (\mathbf{ v_{c_i}}, \mathbf{ v_{c_i}^+}) / \tau ) }}{e^{(sim(\mathbf{ v_{c_i}},\mathbf{ v_{c_i}^+} )/\tau)} + \sum\limits_{k=1}^{K} e^{ (sim (\mathbf{ v_{c_i}}\cdot \mathbf{ v_{c_k}^-} )/ \tau)} } \\
\end{aligned}
\end{equation}
where ${q_i}^+$ is the paired query of input code snippet $c_i$, and ${q_k}^-$ denotes the unpaired query.

The inter-modal and intra-modal loss function in a mini-batch can be obtained by:
\begin{equation}
\small
L^{inter} = \sum_{i=1}^{bs} ( L^{inter}_{q_i} +  L^{inter}_{c_i}), \ L^{intra} = \sum_{i=1}^{bs} ( L^{intra}_{q_i} +  L^{intra}_{c_i}) 
\label{eq:inter-intra}
\end{equation}

To this end, the overall multimodal contrastive learning loss function for a mini-batch is:
\begin{equation}
\small
 L = \sum_{i=1}^{bs} ( L^{inter} +  L^{intra})
\end{equation}
We apply AdamW~\cite{loshchilov2017decoupled} to optimize the overall model.

\subsection{Fine-tuning on Code Search}
After being optimized by multimodal contrastive learning, the model can learn better representations of samples, where similar samples (code or queries) have similar representations and different samples have different representations. To further improve the performance of model on code search, following the most previous studies~\cite{FengGTDFGS0LJZ20,GuoRLFT0ZDSFTDC21,0034WJH21,AhmadCRC21,NiuL0GH022,wang2021syncobert,GuoLDW0022}, we fine-tune it on the related training dataset by:
\begin{equation}
\small
\begin{aligned}
    L^{f} = -\sum_{i=1}^{bs} [ \log \frac{ e^{ (sim (\mathbf{ v_{c_i}}, \mathbf{ v_{q_i}}) / \tau )} }{\sum_{j=1}^{bs} e^{(sim ( \mathbf{ v_{c_i}}\cdot \mathbf{ v_{q_j}} )/ \tau)} } ]
\end{aligned}
\label{eq:in_batch_loss}
\end{equation}
where  $\mathbf{v_{c_i}}$ and $\mathbf{ v_{q_i}}$ are the overall semantic representations of the code snippet $c_i$ and query $q_i$, respectively. They are obtained by code and query encoder, respectively. $\tau$ is the temperature hyperparameter. Then we use the validation dataset to select the best model based on the MRR value (details in \Sec~\ref{sec:metric}) and report scores on the test set in this paper.

\section{Experimental Design}
\label{experimental_design}

\subsection{Datasets}
\begin{table}[t]
\setlength{\tabcolsep}{4pt}
\centering
\tablesize
\caption{Dataset statistics.}  
\begin{tabular}{lcccc} 
\toprule
 Language & \text { Training } & \text { Validation } & \text { Test } & Candidate Code
 \\
\midrule 
\text { \ruby{} } & 24,927 & 1,400 & 1,261 & 4,360 \\
\text { \javascript } & 58,025 & 3,885 & 3,291 & 13,981 \\
\text { \java } & 164,923 & 5,183 & 10,955 & 40,347 \\
\text { \go } & 167,288 & 7,325 & 8,122 & 28,120 \\
\text { \php } & 241,241 & 12,982 & 14,014 & 52,660 \\
\text { \python} & 251,820 & 13,914 & 14,918 & 43,827 \\
\bottomrule
\end{tabular}
\label{tab:dataset}
\end{table}

We conduct experiments on a large-scale benchmark dataset  CodeSearchNet~\cite{Husain2019} as used in Guo et al.~\cite{GuoRLFT0ZDSFTDC21}. It contains six programming languages, namely \ruby, \javascript, \go, \python, \java, and \php. This dataset is widely used in previous studies~\cite{GuoRLFT0ZDSFTDC21,FengGTDFGS0LJZ20,Husain2019,DuSWSHZ21,0034WJH21,wang2020cocogum,GuLGWZXL21,NiuL0GH022,GuoLDW0022}. 
The statistics of the dataset are shown in \Tab~\ref{tab:dataset}. 
Following previous studies~\cite{GuZ018, GuoRLFT0ZDSFTDC21,HuangTSG0J0D20}, the model is to retrieve the correct code snippets from the \emph{Candidate Code} (the last column in \Tab~\ref{tab:dataset}) for the given queries when performing the evaluation.

\subsection{Baselines}
\label{baselines}
To evaluate the effectiveness of our approach, we compare \Our{} with \revise{three IR-based methods~\cite{nbowschutze2008introduction,tfrobertson1976relevance,jaccard1901etude}}, four deep end-to-end approaches~\cite{Husain2019} and ten pre-training-based approaches including three contrastive learning-related models. 

\begin{itemize}[leftmargin=10pt]
    \item \revise{IR-based methods include \textbf{BOW}~\cite{nbowschutze2008introduction}, \textbf{TF-IDF}~\cite{tfrobertson1976relevance} and \textbf{Jaccard}~\cite{jaccard1901etude}. BOW and TF-IDF use bag-of-word and term frequency-inverse document frequency techniques, respectively, to extract the features from the input code snippets and queries and convert them into vectors. Then, they measure semantic similarities between code snippets and queries by cosine similarities between their corresponding vectors. Jaccard retrieves the similar code snippet for the given query according to the Jaccard similarity coefficient~\cite{jaccard1901etude} between the code and query.}   
    \item \textbf{NBow}, \textbf{CNN},  \textbf{BiRNN} and  \textbf{SelfAtnn}~\cite{Husain2019} use various encoding models such as neural bag-of-words~\cite{IyyerMBD15}, 1D convolultional neural network~\cite{Kim14} , bi-directional GRU~\cite{ChoMGBBSB14}, and multi-head attention~\cite{VaswaniSPUJGKP17} to obtain the representations of code snippets and queries. And they measure the semantic similarity of representations using inner product.
    
    \item \textbf{\roberta}~\cite{Liu2019Roberta}, \textbf{\roberta{} (code)}~\cite{FengGTDFGS0LJZ20} are built on a multi-layer Transformer encoder~\cite{VaswaniSPUJGKP17} and pre-trained with MLM, which is to predict the masked tokens. The  pre-trained datasets are natural language corpus ~\cite{Liu2019Roberta} and source code corpus~\cite{Husain2019}, respectively.
    
    \item \textbf{\codebert}~\cite{FengGTDFGS0LJZ20} and \textbf{\graphcodebert}~\cite{GuoRLFT0ZDSFTDC21} are pre-trained on a large code corpus. The formal is pre-trained with MLM and RTD, which uses a discriminator to identify replaced tokens. The latter considers the code structure information and is pre-trained tasks with  MLM, data flow edge prediction, and node alignment.
    
     \item \textbf{\sigircon}~\cite{BuiYJ21} firstly uses a unimodal contrastive learning approach (only the code modality) to pre-train the model to recognize the semantically equivalent code snippets and then fine-tune it on the downstream tasks. \revise{As \sigircon does not release the implementation of semantic-preserving transformations and it is costly to implement these transformations for six programming languages, we only implemented for \java language because \java is the most studied language for code search~\cite{LiuXLGYG22}. To make a fair comparison, we use the unimodal contrastive learning technique to continually pre-train the \unixcoder and then fine-tune it on the code search task as the implementation of \sigircon.}

     \item \revise{\textbf{ContraCode}~\cite{0001JZA0S21} also applies contrastive learning to unsupervised code representation learning and conducts experiments on code summarization and type inference. Specifically, they first adopt some semantic-preserving program transformations to generate functional equivalence code snippets. Next, they pre-train a neural network model to identify functionally equivalent code snippets among many distractors. Finally, they fine-tune the pre-trained model to perform downstream tasks. We use the pre-trained ContraCode as the code/query encoder and optimize it using~\Eq~\ref{eq:in_batch_loss} for code search.}
      
     \item \textbf{\codetf}~\cite{0034WJH21}, \textbf{\plbart}~\cite{AhmadCRC21}, \textbf{\sptcode}~\cite{NiuL0GH022} are sequence-to-sequence code pre-trained models. The first is pre-trained with three identifier-aware pre-training tasks to enable the model to identify identifiers in source code or recover masked identifiers. The second is pre-trained with denoising autoencoding, which is to reconstruct the corrupted input code sequence. The third takes source code,  corresponding AST and paired summarization as input and is pre-trained with three code-specific tasks.
     
    \item \textbf{\syncobert}~\cite{wang2021syncobert} and \textbf{\unixcoder}~\cite{GuoLDW0022} are multi-modal contrastive pre-training for code representation. \syncobert takes source code, AST and summarization as input and pre-trained with identifier prediction and AST edge prediction to learn the lexical and syntactic knowledge of source code. \unixcoder{} takes two-modality data, the summarization and simplified AST of source code, as input and is pre-trained with MLM, unidirectional language modeling, denoising autoencoder, and two contrastive learning-related tasks.
    
\end{itemize}

More details about baselines can be found in Appendix of the replication package~\cite{anonymousRepo}. In our experiments, we train the four deep end-to-end approaches from scratch, and for the ten pre-trained approaches, we initialize them with the pre-trained models and fine-tune (or continually pre-train and fine-tune) them according to the descriptions in their original papers or their released source code.

\subsection{Experimental Settings}
\label{exp_setting}
Following \unixcoder~\cite{GuoLDW0022}, we use Transformer with 12 layers, 768 dimensional hidden states, and 12 attention heads. The vocabulary sizes of code and queries are set to 51,451. Max sequence lengths of code snippets and queries are 128 and 256, respectively. 
For optimizer, we use AdamW with the learning rate 2e-5.  
Following previous studies~\cite{FengGTDFGS0LJZ20,GuoRLFT0ZDSFTDC21,HuangTSG0J0D20}, the code encoder and query encoder share parameters to reduce the number of total parameters. Following MoCo~\cite{He0WXG20}, the temperature hyperparameter $\tau$ is set as 0.07 and momentum coefficient $m$ is 0.999. The queue size and batch size are set to 4096 and 128, respectively. The training step of multimodal contrastive learning stage is 100K and the maximum epochs of fine-tune stage is 5. In addition, we run the experiments 3 times with random seeds 0,1,2 and display the mean value in the paper.
All experiments are conducted on a machine with 220 GB main memory and Tesla A100 80GB GPU.

\subsection{Evaluation Metrics}
\label{sec:metric}
We measure the performance of our approach using four metrics: mean reciprocal rank (MRR) and  top-k recall (R@k, k=1,5,10), which are widely used in previous studies~\cite{GuoRLFT0ZDSFTDC21,FengGTDFGS0LJZ20,Husain2019,GuZ018,GuCM21,HaldarWXH20,ShuaiX0Y0L20,WanSSXZ0Y19,YeXZHWZ20,LiQYSC20,LingLZX20,LingWWPMXLWJ21,WanSSXZ0Y19,WanSSXZ0Y19,DuSWSHZ21,ZhuSLXZ20}. \textbf{MRR} is the average of reciprocal ranks of the correct code snippets for given queries $Q$. \textbf{R@k} measures the percentage of queries that the paired code snippets
exist in the top-k returned ranked lists. They are calculated as follows:
\begin{equation}
\small
   MRR=\frac{1}{|Q|} \sum_{i=1}^{|Q|} \frac{1}{Rank_i}, \  \text { R@k }=\frac{1}{|Q|} \sum_{i=1}^{|Q|} \delta\left(Rank_i \leq k\right)
\end{equation}
\noindent where $Rank_i$ is the rank of the paired code snippet related to the \textit{i}-th query. $\delta$ is an indicator function that returns 1 if $Rank_i \leq k$ otherwise returns 0.

\section{Experimental Results}
\label{exprimental_result}

\subsection{RQ1: What Is the Effectiveness of \Our{}?}
\label{RQ1}
\subsubsection{Overall results}

\begin{table}[t]
 \centering
\tablesize
\setlength{\tabcolsep}{1.0pt}
\caption{Performance of different approaches. JS is short for \javascript. Statistical significance of experiments: $p <0.01$.} 
\begin{tabular}{llccccccc} 
\toprule
&Model & \ruby{} & JS &\go  & \python & \java & \php & \avg \\
\midrule
\multirow{3}{*}{\shortstack{IR-Based \\models}}
& BOW&0.230&0.184&0.350&0.222&0.245&0.193&0.237\\
& TF-IDF&0.239&0.204&0.363&0.240&0.262&0.215&0.254\\
& Jaccard&0.220&0.191&0.345&0.243&0.235&0.182&0.236\\
\midrule
\multirow{4}{*}{\shortstack{Deep \\ end-to-end \\models}}
&\nbow & 0.162 & 0.157 & 0.330 & 0.161 & 0.171 & 0.152 & 0.189 \\
&\cnn & 0.276 & 0.224 & 0.680 & 0.242 & 0.263 & 0.260 & 0.324 \\
&\birnn & 0.213 & 0.193 & 0.688 & 0.290 & 0.304 & 0.338 & 0.338 \\
&\selfann & 0.275 & 0.287 & 0.723 & 0.398 & 0.404 & 0.426 & 0.419 \\

\midrule
\multirow{10}{*}{\shortstack{Pre-trained  \\ models}}
&\roberta &0.587 &0.523 &0.855 &0.590 &0.605 &0.561 &0.620 \\
&\roberta (code)  & 0.631 &0.57 &0.864 &0.621 &0.636 &0.581 &0.650 \\
&\codebert &0.679 &0.621 &0.885 &0.672 &0.677 &0.626 &0.693\\
&\graphcodebert  &0.703 &0.644 & 0.897  & 0.692  &0.691 & 0.649  & 0.713 \\
&\sigircon &- &- &- &-&0.727  &-  &- \\
&ContraCode &- &0.688 &- &-&-  &-  &- \\
&\plbart     &0.675  &0.616 &0.887 &0.663 &0.663 &0.611 &0.685\\
&\codetf &0.719  &0.655 &0.888 &0.698 &0.686 &0.645 &0.715\\
&\syncobert   &0.722  &0.677 &0.913 &0.724 &0.723 &0.678 &0.740\\
&\unixcoder   &0.740  &0.684 &0.915 &0.720 &0.726 &0.676 &0.744\\
&\sptcode    &0.701  &0.641 &0.895 &0.699 &0.700 &0.651 &0.715\\
\midrule
\multirow{2}{*}{Our} 
  &\multirow{2}{*}{\Our{}}  &\textbf{0.818} &\textbf{0.764} &\textbf{0.921}  &\textbf{0.757} &\textbf{0.763} &\textbf{0.703} &\textbf{0.788} \\
 &&$\uparrow$10.54\% &$\uparrow$11.7\% &$\uparrow$0.66\% &$\uparrow$5.14\% &$\uparrow$5.10\% &$\uparrow$3.99\% &$\uparrow$5.91\%\\
\bottomrule
\end{tabular}
\label{tab:comp_with_baselines}
\end{table}

We evaluate the effectiveness of our model \Our{} by comparing it to three IR-based models, four recent deep end-to-end code search models and ten pre-trained models introduced in \Sec{}~\ref{baselines} on the CodeSearchNet dataset with six programming languages. As there is no released source code for \sigircon and it is costly to reproduce this approach for six programming languages, we reproduce \sigircon on \java language for comparison. The experimental results are shown in \Tab~\ref{tab:comp_with_baselines}. We present the results under MRR metric only due to space limitation. We put results under other metrics in Appendix of the replication package~\cite{anonymousRepo}. Conclusions that hold on MRR also hold for other metrics.

We can see that~\revise{deep end-to-end models outperform IR-based models because they can learn better semantic relations between codes and queries~\cite{GuZ018,LiQYSC20,WanSSXZ0Y19}.} The ten pre-trained models (the third row of \Tab~\ref{tab:comp_with_baselines}) perform better than the four deep end-to-end models (the second row of \Tab~\ref{tab:comp_with_baselines}) trained from scratch, which shows the effectiveness of the pre-training technique. Since \graphcodebert considers the data flow information of source code, it performs better than \roberta, \roberta (code) and \codebert. \unixcoder and \syncobert consider the structure information of source code and perform better than other approaches on average.
\revise{\sigircon and ContraCode perform best among baselines because they pre-train the model to recognize the functionally equivalent code snippets from many distractors and can learn a better code representation.} \Our{} takes \unixcoder as the base code/query encoder, continually optimizes it with multimodal contrastive learning and soft data augmentation and performs best among all approaches.

\subsubsection{Case study}
Next, we show some cases to demonstrate the effectiveness of our model \Our{}. \revise{For each case, we only show the result of our approach and the best baseline, which is GraphCodeBERT for the 1st case and UniXcoder for the 2nd case.} 

\Fig~\ref{fig:case1} shows the results returned by \Our{} and \graphcodebert{} for the query ``Transform a hexadecimal String to a byte array.'' from the Java dataset. The query includes two operation objects: ``hexadecimal String'' and ``byte array'' and one action ``Transform''. To implement the functionality of the query, we usually take the ``hexadecimal String'' as an input parameter and use ``toXXX(...)'' to perform the ``Transform'' action. Our model \Our{} can successfully understand the semantics of the whole query and code snippet and return the correct result, while \graphcodebert cannot. This is because the code representation obtained by \graphcodebert is affected by token-level semantics such as \texttt{String} and \texttt{byte}, 
thereby, returning the code snippet which has many similar tokens with the query but a completely opposite semantic: \textit{``Converts bytes to hex string."}.

\begin{figure}[t]
\begin{lstlisting}[language=java,caption={The top-1 result  returned by \Our{}.}]
public static byte[] hexStringToByte(String hexString) {
    try {
        return Hex.decodeHex(hexString.toCharArray());} 
    catch (DecoderException e) {
        throw new UnexpectedException(e);}}
\end{lstlisting}
\vspace{5pt}

\begin{lstlisting}[language=java,caption={The top-1 result returned by \graphcodebert{}.},]
public static String toHexString(final byte[] bytes) {
    char[] chars = new char[bytes.length * 2];
    int i = 0;
    for (byte b : bytes) {
        chars[i++] = CharUtil.int2hex((b & 0xF0) >> 4);
        chars[i++] = CharUtil.int2hex(b & 0x0F);}
    return new String(chars);}
\end{lstlisting}
\caption{The top-1 code returned by \Our{} and \graphcodebert for the query  ``Transform a hexadecimal String to a byte array.'' on \java language.}
\label{fig:case1}
\vspace{-2.5mm}
\end{figure}

\begin{figure}[t]
\begin{lstlisting}[language=python,caption={The top-1 result returned by \Our{}.},]
def get_items(self):
    reader = csv.reader(self.source)
    headers = reader.next()
    for row in reader:
        if not row:
            continue
        yield dict(zip(headers, row))
\end{lstlisting}
\vspace{2pt}

\begin{lstlisting}[language=python,caption={The top-1 result returned by \unixcoder{}.}]
def iterrows(lines_or_file, namedtuples=False, dicts=False, encoding='utf-8', **kw):
    if namedtuples and dicts:
        raise ValueError('either namedtuples or dicts can be chosen as output format')
    elif namedtuples:
        _reader = NamedTupleReader
    elif dicts:
        _reader = UnicodeDictReader
    else:
        _reader = UnicodeReader
    with _reader(lines_or_file, encoding=encoding, **fix_kw(kw)) as r:
        for item in r:
            yield item
\end{lstlisting}
\caption{The top-1 code returned by \Our{} and \unixcoder for the query  ``Iterator to read the rows of the CSV file.'' on \python language.}
\label{fig:case2}
\vspace{-2.5mm}
\end{figure}

In \Fig~\ref{fig:case2}, we compare the results returned by \Our{} and \unixcoder{} for the query ``Iterator to read the rows of the CSV file.''  in the \python dataset. \Our{} returns the correct code snippet, while \unixcoder returns the code snippet with another semantics ``Yield a generator over the rows.''. Our model can accurately understand the intent of the query and return the relevant code snippet, which first reads a CSV file and yields an iterator to read the rows of it. \unixcoder{} returns a partially relevant code snippet which only yields an iterator to read the rows of files, missing reading a CSV file.

\begin{tcolorbox}
\textbf{Summary.} Our approach significantly outperforms baselines on six programming languages in terms of four metrics. Case studies further demonstrate the advantages of \Our{} in code search.
\end{tcolorbox}
 
\subsection{RQ2: How Much Do Different Components Contribute?}
\label{RQ2:ablation-study}
\begin{table}[t]
 \centering
\tablesize
\setlength{\tabcolsep}{5pt}
\caption{Ablation study of \Our{} on MRR. }
\begin{tabular}{lccccccc} 
\toprule
Model  & \ruby{} & JS &\go  & \python & \java & \php & \avg \\
\midrule
\textbf{\Our{}}  &\textbf{0.818} &\textbf{0.764} &\textbf{0.921}  &\textbf{0.757} &\textbf{0.763} &\textbf{0.703} &\textbf{0.788}  \\ 
\ w/o \ShortReplacement &0.794 &0.714 &0.905 &0.730 &0.743 &0.688 &0.762 \\
\ w/o \ShortMasking &0.801 &0.754 &0.906 &0.744 &0.756 &0.69 &0.775 \\
\ w/o \ShortTypeReplacement &0.797 &0.714 &0.905 &0.730 &0.743 &0.685 &0.762 \\
\ w/o \ShortTypeMasking &0.788 &0.715 &0.902 &0.731 &0.746 &0.686 &0.761 \\
\ w/o all \soda{}s &0.775 &0.711 &0.907 &0.736 &0.738 &0.683 &0.758 \\
\ w/o inter-modal loss &0.776 &0.703 &0.906 &0.729 &0.739 &0.674 &0.755 \\
\ w/o intra-modal loss &0.780 &0.705 &0.903 &0.727 &0.744 &0.680 &0.756 \\
\bottomrule
\end{tabular}
\label{tab:ablation}
\end{table}

In this section, we study the contribution of each component of our approach \Our{}. It  includes multimodal contrastive learning such as intra-modal loss and inter-model loss and four \soda approaches (\ShortReplacement, \ShortMasking, \ShortTypeReplacement and \ShortTypeMasking) introduced in \Sec~\ref{sec:soda}. Specifically, we remove one component (such as \ShortMasking) of \Our{} each time and then study the performance of the ablated model. The experimental results are shown in~\Tab~\ref{tab:ablation}. ``w/o one component" means to remove this component. For example, \Our{} w/o \ShortReplacement and  inter-modal loss means to drop the \soda method \ShortReplacement and inter-modal loss function $L^{intra}$ (\Eq~\ref{eq:inter-intra}), respectively.

From the \Tab~\ref{tab:ablation}, we can see that the performance of the model drops after removing any one component. It demonstrates that each component plays an important role in the code search model. Especially, the performance of \Our{} w/o all \soda{}s, intra-modal and inter-modal contrastive learning loss drops obviously. This is because data augmentation can increase data diversity.
The inter-modal contrastive learning, which aims to learn the alignment of code snippets and queries, can pull together the paired code and query and push apart the unpaired code and query. Intra-modal contrastive learning aims to learn a uniform distribution of representation of unimodal samples (code snippets or queries) and can improve the generalization performance of code search model.

\begin{tcolorbox}
\textbf{Summary.}  The ablation study shows the effectiveness of multimodal contrastive learning including intra-modal loss and inter-model loss and four \soda approaches including \ShortReplacement, \ShortMasking, \ShortTypeReplacement and \ShortTypeMasking.
\end{tcolorbox}

\subsection{RQ3: What Is the Performance of Our Approach on Other Pre-trained Models?}
\label{RQ_other_pre_train}

\begin{table*}[t]
 \centering
\tablesize
\setlength{\tabcolsep}{5pt}
\caption{Results on other pre-trained models. \graphcodebertOurShort is short for \graphcodebertOur. The improved percentages are shown in parentheses.}
\begin{NiceTabular}{lp{1cm}|cc|cc|cc} 
\toprule

PL & Metric & \roberta & \robertaOur  &\codebert &\codebertOur  &\graphcodebert &\graphcodebertOurShort   \\

\midrule
\multirow{4}{*}{\ruby{}}
&MRR &0.587 &\textbf{0.640} ($\uparrow$9.03\%) &0.679 &\textbf{0.723} ($\uparrow$6.48\%) &0.703 &\textbf{0.752} ($\uparrow$6.97\%)\\
&R@1 &0.469 &\textbf{0.533} ($\uparrow$13.65\%) &0.583 &\textbf{0.618} ($\uparrow$6.00\%) &0.607 &\textbf{0.655} ($\uparrow$7.91\%)\\
&R@5 &0.717 &\textbf{0.764} ($\uparrow$6.56\%) &0.800 &\textbf{0.852} ($\uparrow$6.50\%) &0.824 &\textbf{0.875} ($\uparrow$6.19\%)\\
&R@10 &0.785 &\textbf{0.825} ($\uparrow$5.10\%) &0.853 &\textbf{0.904} ($\uparrow$5.98\%) &0.872 &\textbf{0.916} ($\uparrow$5.05\%)\\
\hline
 \multirow{4}{*}{\javascript{}}
&MRR &0.523 &\textbf{0.559} ($\uparrow$6.88\%) &0.621 &\textbf{0.648} ($\uparrow$4.35\%) &0.644 &\textbf{0.682} ($\uparrow$5.90\%)\\
&R@1 &0.413 &\textbf{0.460} ($\uparrow$11.38\%) &0.514 &\textbf{0.545} ($\uparrow$6.03\%) &0.538 &\textbf{0.582} ($\uparrow$8.18\%)\\
&R@5 &0.652 &\textbf{0.673} ($\uparrow$3.22\%) &0.752 &\textbf{0.772} ($\uparrow$2.66\%) &0.774 &\textbf{0.806} ($\uparrow$4.13\%)\\
&R@10 &0.730 &\textbf{0.744} ($\uparrow$1.92\%) &0.814 &\textbf{0.839} ($\uparrow$3.07\%) &0.834 &\textbf{0.866} ($\uparrow$3.84\%)\\
\hline
 \multirow{4}{*}{\go{}}
&MRR &0.855 &\textbf{0.881} ($\uparrow$3.04\%) &0.885 &\textbf{0.905} ($\uparrow$2.26\%) &0.897 &\textbf{0.907} ($\uparrow$1.11\%)\\
&R@1 &0.800 &\textbf{0.829} ($\uparrow$3.62\%) &0.837 &\textbf{0.859} ($\uparrow$2.63\%) &0.858 &\textbf{0.861} ($\uparrow$0.35\%)\\
&R@5 &0.926 &\textbf{0.945} ($\uparrow$2.05\%) &0.944 &\textbf{0.962} ($\uparrow$1.91\%) &0.954 &\textbf{0.962} ($\uparrow$0.84\%)\\
&R@10 &0.949 &\textbf{0.965} ($\uparrow$1.69\%) &0.962 &\textbf{0.975} ($\uparrow$1.35\%) &0.972 &\textbf{0.978} ($\uparrow$0.62\%)\\
\hline
 \multirow{4}{*}{\python{}}
&MRR &0.590 &\textbf{0.629} ($\uparrow$6.61\%) &0.672 &\textbf{0.690} ($\uparrow$2.68\%) &0.692 &\textbf{0.714} ($\uparrow$3.18\%)\\
&R@1 &0.480 &\textbf{0.523} ($\uparrow$8.96\%) &0.574 &\textbf{0.589} ($\uparrow$2.61\%) &0.594 &\textbf{0.614} ($\uparrow$3.37\%)\\
&R@5 &0.727 &\textbf{0.756} ($\uparrow$3.99\%) &0.792 &\textbf{0.809} ($\uparrow$2.15\%) &0.813 &\textbf{0.834} ($\uparrow$2.58\%)\\
&R@10 &0.793 &\textbf{0.818} ($\uparrow$3.15\%) &0.850 &\textbf{0.867} ($\uparrow$2.00\%) &0.866 &\textbf{0.888} ($\uparrow$2.54\%)\\
\hline
 \multirow{4}{*}{\java{}}
&MRR &0.605 &\textbf{0.635} ($\uparrow$4.96\%) &0.677 &\textbf{0.705} ($\uparrow$4.14\%) &0.691 &\textbf{0.721} ($\uparrow$4.34\%)\\
&R@1 &0.499 &\textbf{0.531} ($\uparrow$6.41\%) &0.580 &\textbf{0.606} ($\uparrow$4.48\%) &0.592 &\textbf{0.624} ($\uparrow$5.41\%)\\
&R@5 &0.737 &\textbf{0.762} ($\uparrow$3.39\%) &0.796 &\textbf{0.826} ($\uparrow$3.77\%) &0.817 &\textbf{0.843} ($\uparrow$3.18\%)\\
&R@10 &0.796 &\textbf{0.820} ($\uparrow$3.02\%) &0.852 &\textbf{0.878} ($\uparrow$3.05\%) &0.865 &\textbf{0.890} ($\uparrow$2.89\%)\\
\hline
 \multirow{4}{*}{\php{}}
&MRR &0.561 &\textbf{0.598} ($\uparrow$6.60\%) &0.626 &\textbf{0.647} ($\uparrow$3.35\%) &0.649 &\textbf{0.668} ($\uparrow$2.93\%)\\
&R@1 &0.450 &\textbf{0.490} ($\uparrow$8.89\%) &0.520 &\textbf{0.538} ($\uparrow$3.46\%) &0.545 &\textbf{0.561} ($\uparrow$2.94\%)\\
&R@5 &0.694 &\textbf{0.730} ($\uparrow$5.19\%) &0.753 &\textbf{0.779} ($\uparrow$3.45\%) &0.785 &\textbf{0.798} ($\uparrow$1.66\%)\\
&R@10 &0.764 &\textbf{0.797} ($\uparrow$4.32\%) &0.814 &\textbf{0.844} ($\uparrow$3.69\%) &0.832 &\textbf{0.863} ($\uparrow$3.73\%)\\
\bottomrule
\end{NiceTabular}
\label{tab:diff_pre-train_mrr_r}
\end{table*}

We further study the performance of our approach on other three pre-trained models introduced in \Sec{}~\ref{baselines}, including a natural language pre-trained model \roberta and two source code pre-trained models \codebert and \graphcodebert.
Specifically, we use these pre-trained models as the code/query encoders and momentum code/query encoders in \Fig~\ref{fig:framework}. For the input code snippet and query sequence, we average hidden states of the last layer as the overall representations of the code snippet or query. The similarities of representations are measured by the cosine similarity. Other experimental settings are same as in ~\Sec{}~\ref{exp_setting}. 

The results are shown in \Tab~\ref{tab:diff_pre-train_mrr_r}.
\robertaOur means using \roberta as the code/query encoders and momentum code/query encoders in our framework.
Overall, we can see that \robertaOur, \codebertOur and \graphcodebertOur obviously outperform \roberta, \codebert and \graphcodebert, respectively on all six programming languages in terms of four metrics. These results demonstrate that our approach can be generalized to other pre-trained models and boost their performance. Besides, \robertaOur, which is pre-trained on the natural language corpus and fine-tuned with our method, achieves comparable performance with \codebert on \go dataset. \graphcodebertOur achieves comparable performance with \unixcoder on \javascript,  \java and \python.
\graphcodebertOur also slightly outperforms  \unixcoder on \ruby  dataset. 

\begin{tcolorbox}
\textbf{Summary.} Our approach is orthogonal to the pre-trained technique on the performance improvement for code search tasks and can obviously boost the performance of existing pre-trained models.
\end{tcolorbox}

\subsection{RQ4: What Is the Impact of Different Hyperparameters?}
\label{RQ_exp_hyperparameter}

\begin{figure}[t]
    \centering
    \begin{minipage}[t]{0.490\linewidth}
        \centering
        \includegraphics[width=1\linewidth]{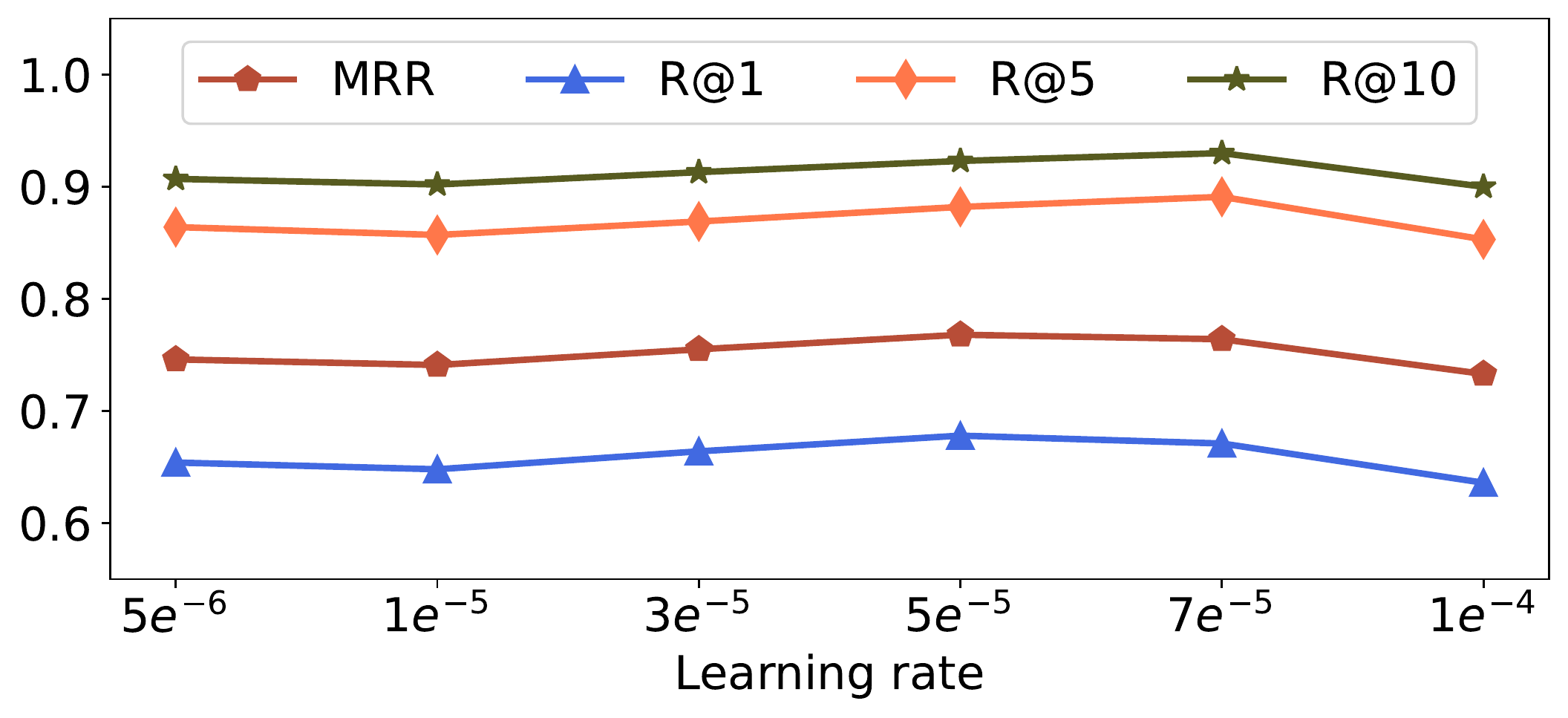}
    \end{minipage}
    \begin{minipage}[t]{0.490\linewidth}
        \centering
        \includegraphics[width=1\linewidth]{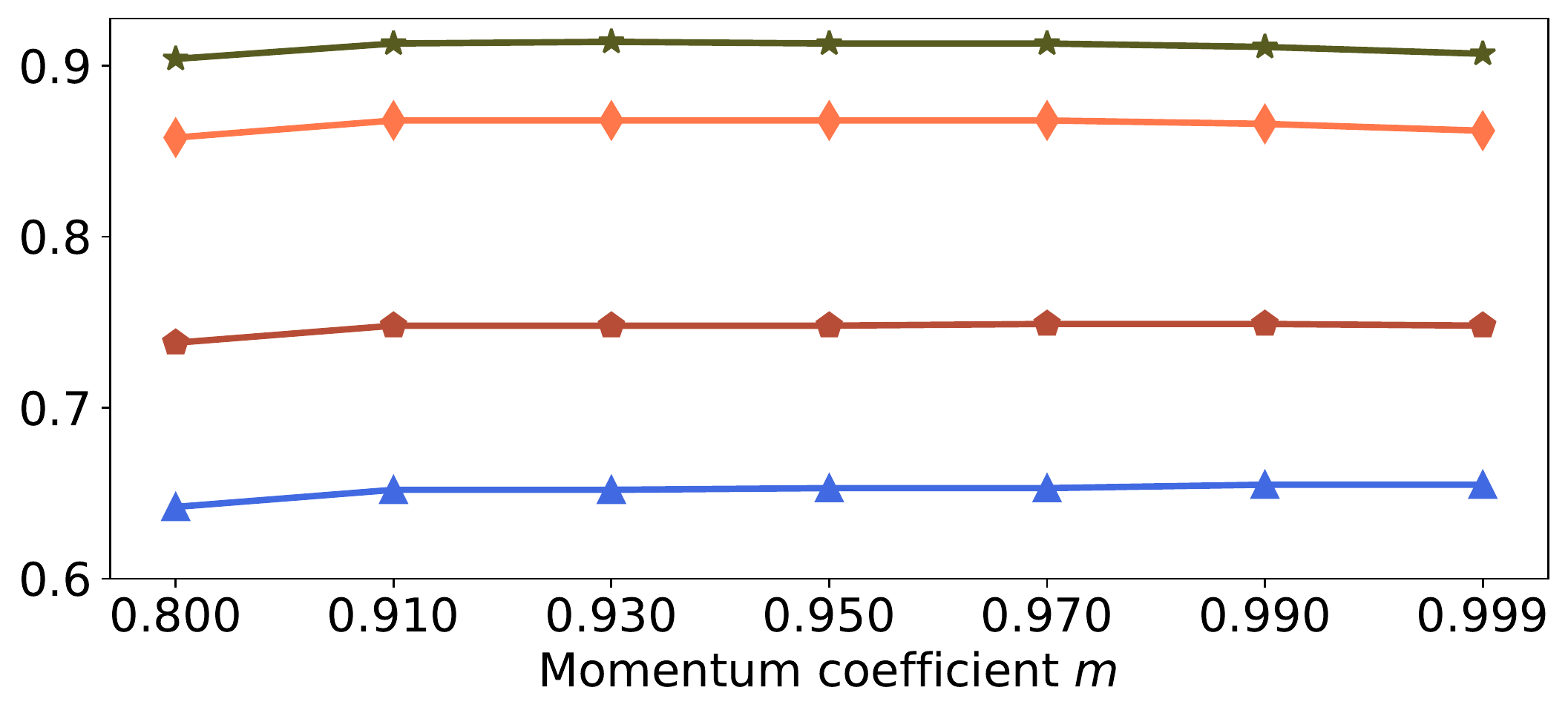}
    \end{minipage}

    \begin{minipage}[t]{0.490\linewidth}
        \centering
        \includegraphics[width=1\linewidth]{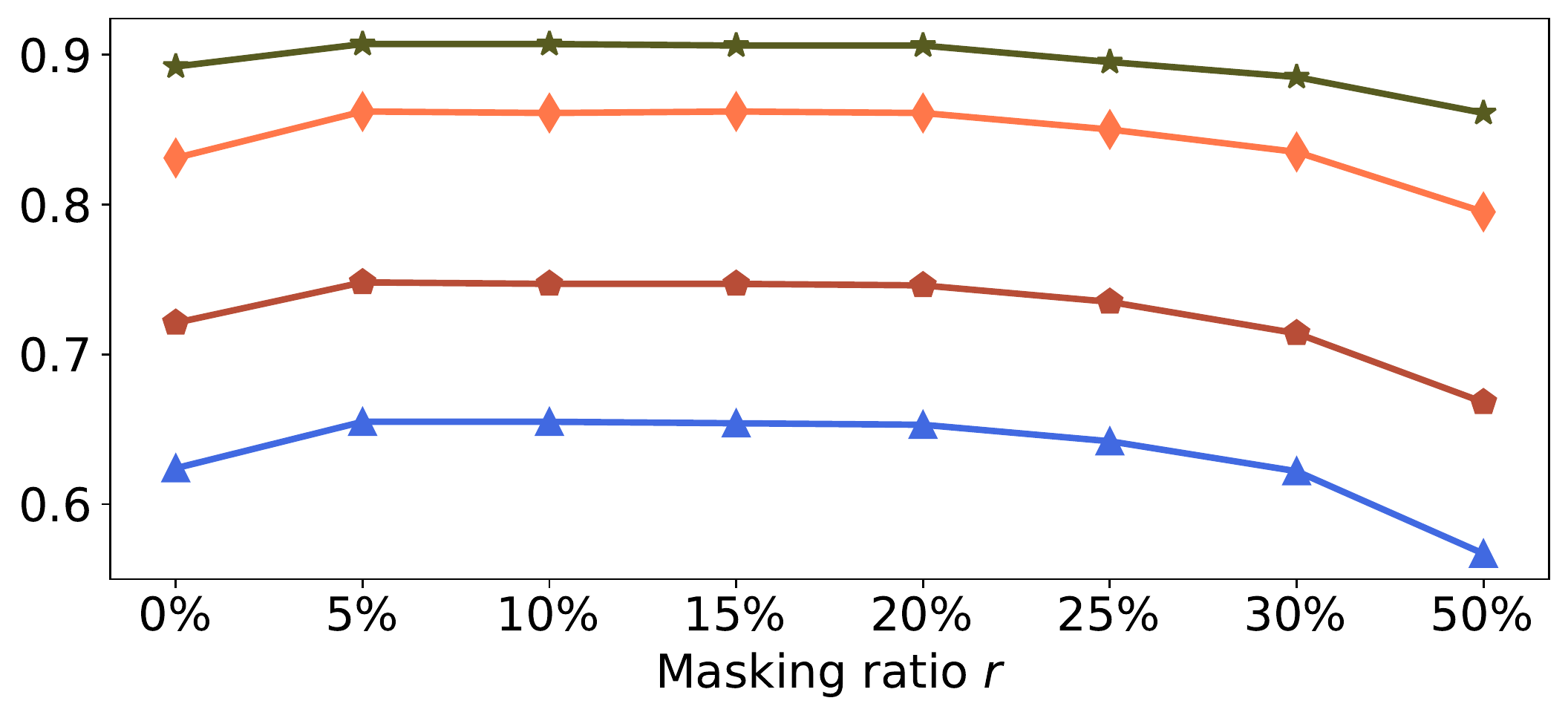}
    \end{minipage}
     \label{fig:diff_r}
        \begin{minipage}[t]{0.490\linewidth}
        \centering 
        \includegraphics[width=1\linewidth]{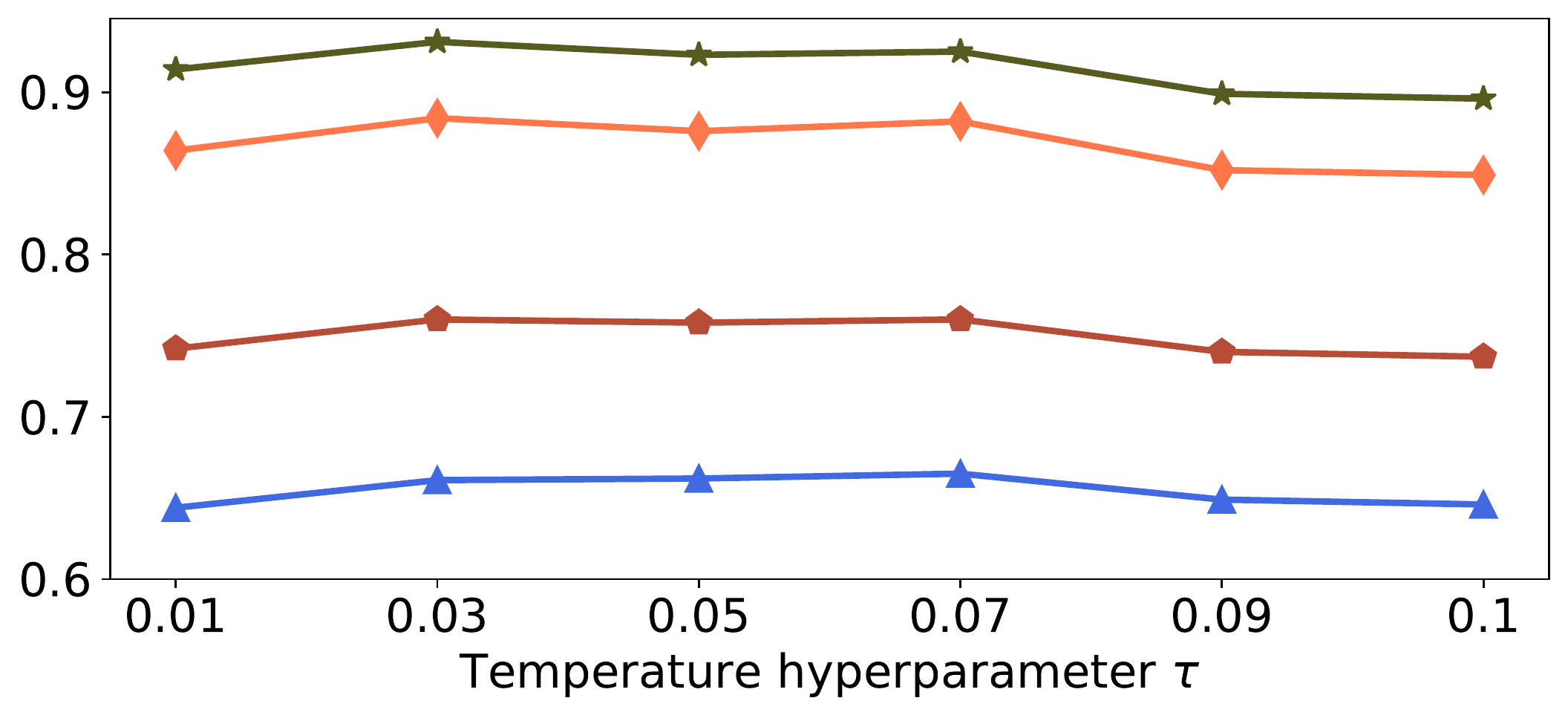}
    \end{minipage}
     \label{fig:diff_r}   
\caption{The impact of different hyperparameters.}
\label{fig:diff_params}
\end{figure}

In this section, we study the impact of different hyperparameters: learning rate, momentum coefficient $m$, masking ratio $r$, and temperature hyperparameter $\tau$.
~\revise{We study different hyperparameters in the typical range, which covers all experimental settings of previous studies~\cite{Liu2019Roberta,FengGTDFGS0LJZ20,GuoRLFT0ZDSFTDC21,BuiYJ21,0034WJH21,NiuL0GH022,wang2021syncobert,GuoLDW0022}} and the experimental results are shown in \Fig~\ref{fig:diff_params}. 
From the results of varying learning rate (the top left of \Fig~\ref{fig:diff_params}),  we can see that performance is generally stable for small learning rate~\cite{MosbachAK21} (from $5e^{-6}$ to $7e^{-5}$). The learning rates that are larger than $7e^{-5}$ have obvious impacts on the model performance. The results of different momentum coefficient $m$ are shown in the top right of \Fig~\ref{fig:diff_params}. We can see that performance increases when the momentum coefficient $m$ becomes larger. This is because a large momentum coefficient is beneficial to obtain the consistent representation for the queue~\cite{He0WXG20}.
The momentum coefficient that is smaller than 0.910 has a significant impact on performance. These findings are consistent with the previous work~\cite{He0WXG20}. From the results of varying masked ratio $r$ (the bottom left of \Fig~\ref{fig:diff_params}), we can see that the performance is insensitive to the masked ratio $r$ when the masked ratio $r$ is between 5\% and 20\%. A larger masked ratio such as 50\% brings considerable performance degradation. It is reasonable because the larger masked ratio causes the code snippet to lose too much information. The results of different temperature hyperparameter $\tau$ are shown in the bottom right of \Fig~\ref{fig:diff_params}. We can see that performance is stable when the temperature hyperparameter $\tau$ varies from 0.03 to 0.07.

\begin{tcolorbox}
\textbf{Summary.} In general, our model is stable over a range of hyperparameter values (learning rate is from $5e^{-6}$ to $7e^{-5}$, momentum coefficient is between 0.910 and 0.999, masked ratio $r$ is from 5\% to 20\%, and temperature hyperparameter $\tau$ varies from 0.03 to 0.07).
\end{tcolorbox}

\subsection{RQ5: Why Does Our Model \Our{} Work?}
The advantages of \Our{} mainly come from soft data augmentation and multimodal momentum contrastive learning. 
The soft data augmentation transforms the input sequence with the masking or replacement mechanism and generates a similar sample as the ``positive'' example. It can help the model learn a representation of the code snippet and query from a global (sequence-level) view rather than simply aggregate the token-level semantic. Thus, our model tends to return code snippets related to the sequence-level functionality rather than the token-level similarities. 
The momentum mechanism enlarges negative samples, allowing to distinguish one sample from more negative samples at each iteration. For multimodal contrastive learning, the inter-modal contrastive learning can pull together the representations of the code-query pair and push apart the representations of queries and many unpaired code snippets, while intra-modal contrastive learning can learn a uniform distribution of representations in terms of unimodal data (code snippets or queries). Therefore, \Our{} can learn better representations of code and queries and perform well on code search. We further explore why \Our{} works through quantitive and qualitative analysis.

\subsubsection{Quantitive analysis}

We explore to understand reasons behind good performance of our approach through  $\ell$\textsubscript{align} and $\ell$\textsubscript{uniform}~\cite{wangicml2020}, which are usually used as indicators to reflect the quality of representation learned by contrastive learning techniques~\cite{wangicml2020,GaoYC21,WangL21a,MengXBTBHS21}.
Mathematically, they are defined as:
\begin{equation}
\small
\label{eq:align-uniform}
    \begin{aligned}
    &\ell\textsubscript{align} = \mathop{\mathbb{E}}\limits_{(x,y) \sim D_{paired}} \Vert f(x) - f(y) \Vert_2^{2} \\
    &\ell\textsubscript{uniform} = \log \mathop{\mathbb{E}}\limits_{(x,y) \stackrel{\text{i.i.d}}{\sim} D} [e ^{-2\Vert f(x) - f(y) \Vert_2^{2}}] \\
    \end{aligned}
\end{equation}
\noindent where $(x,y) \sim D_{paired}$ means the $x$ and $y$ are paired samples, while $(x,y) \stackrel{\text{i.i.d}}{\sim} D$ mean that $x$ and $y$ are independent identically distributed. $f(x)$ and $f(y)$ are learned representations and $\Vert f(x) - f(y) \Vert_2^{2}$ represent the 2-norm of the distance between them. 
From the \Eq~\ref{eq:align-uniform}, we can know that  $\ell\textsubscript{align}$ always is a non-negative number, while $\ell\textsubscript{uniform}$ is a non-positive number.
In terms of our task,  $\ell\textsubscript{align}$ reflects the degree of alignment of representations of code-query pairs. The closer distances of the paired code snippets and queries are,  the smaller the value (close to zero) of $\ell\textsubscript{align}$ is.  $\ell\textsubscript{uniform}$ reflects the uniformity of the distribution of representations of all code snippets or queries. The more uniform the distribution is,  the smaller the value (close to negative infinity) of $\ell\textsubscript{align}$ is. 
In the extreme case where all representations of code snippets and queries are the same,  both values of $\ell\textsubscript{align}$ and  $\ell\textsubscript{uniform}$ are zero. That is to say, learned representations have perfect alignment but extremely poor uniformity. 
In fact, a model which can learn representations with both better alignment (lower $\ell\textsubscript{align}$ ) and uniformity (lower $\ell\textsubscript{uniform}$) can generally achieve better performance~\cite{wangicml2020,GaoYC21}.

\begin{figure}[t]
    \centering
    \includegraphics[width=0.95\linewidth]{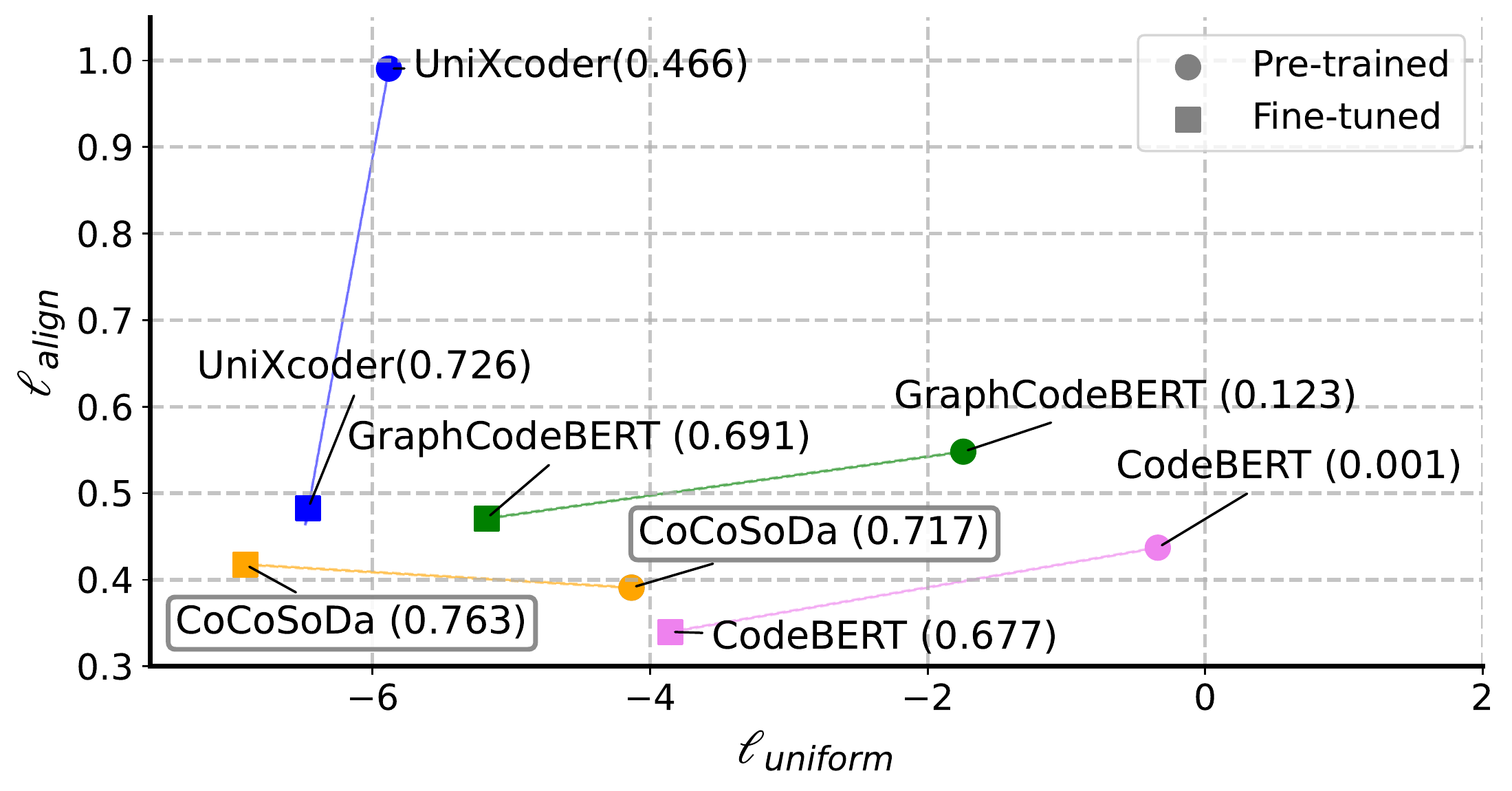} 
      \caption{$\ell$\textsubscript{align}\text{-}$\ell$\textsubscript{uniform} plot of different models. Circles and squares represent pre-trained and fine-tuned models, respectively. Identical models are marked with the same color. MRR scores are shown in parentheses.
      } 
    \label{fig:align_uniform}
\end{figure}

We show $\ell$\textsubscript{align}\text{-}$\ell$\textsubscript{uniform} plot in \Fig~\ref{fig:align_uniform}. 
We can find that \textit{(1)} pre-trained \codebert and \graphcodebert have better alignment but poor uniformity, and perform not well. This is because representations learned by them are very similar and cannot generalize well. \unixcoder has better uniformity and performs better than them. Pre-trained \Our{} has both better alignment and uniformity than~\codebert and~\graphcodebert{} and performs best among four pre-trained models. \textit{(2)} After being fine-tuned, \unixcoder, \codebert and \graphcodebert all have lower $\ell\textsubscript{uniform}$ and $\ell\textsubscript{align}$ and significantly outperform pre-trained ones individually. Our fine-tuned \Our{} generally preserves the alignment and has a better uniform. Therefore, fine-tuned \Our{} further improves the performance of pre-trained it.

\subsubsection{Qualitative analysis}
\begin{figure}[t]
    \centering
    \subfigure[\unixcoder]{
    \begin{minipage}{0.480\linewidth}
        \centering
        \includegraphics[width=1\linewidth]{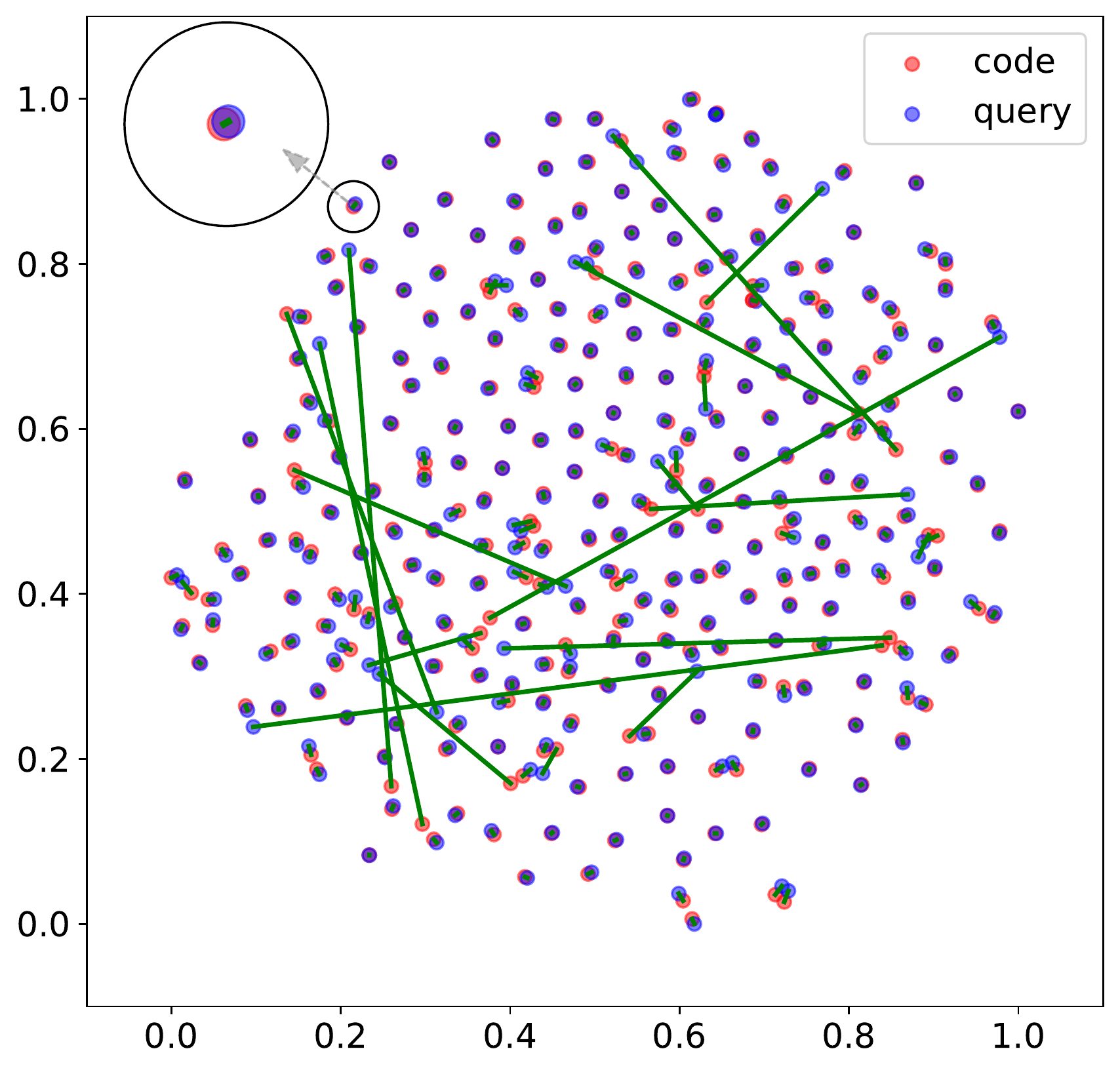}
        \label{fig:t-sne-unixcoder}
    \end{minipage}}
     \subfigure[\Our{}]{
    \begin{minipage}{0.480\linewidth}
        \centering
        \includegraphics[width=1\linewidth]{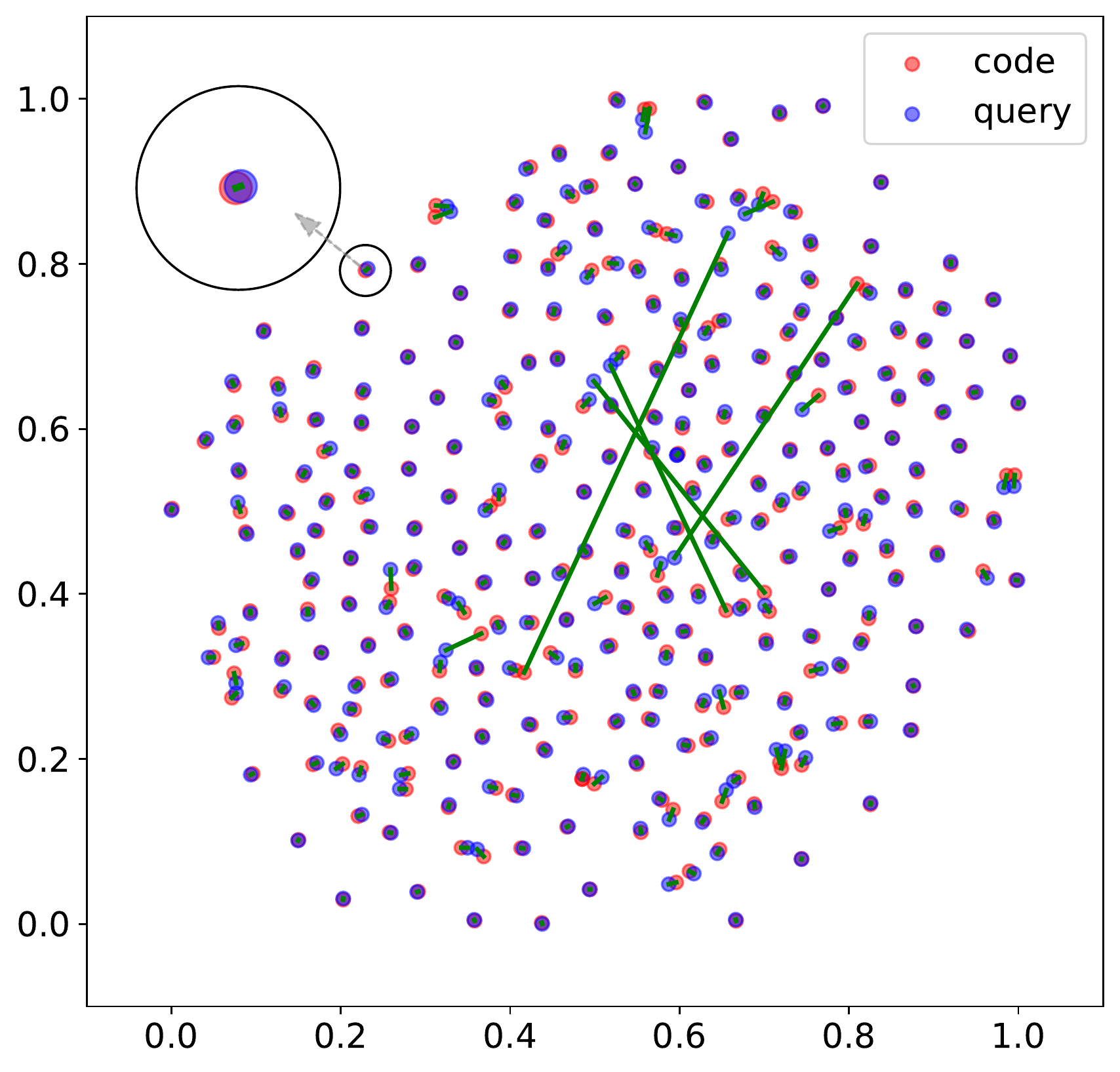}
        \label{fig:t-sne-cocosoda}
    \end{minipage}}
    
\caption{T-SNE visualization of representations of code snippets and queries. (a) and (b) are the representations learned by \unixcoder and \Our{}, respectively. Code is in orange and query is in blue. The distance of paired code and query is indicated by a green line.}
\label{fig:t-sne}
\end{figure}

We also visualize learned representations to help intuitively understand why \Our{} works well. Specifically, first, we randomly sample $X$ ($X$=100, 200, 300, or 400) code-query pairs from \java dataset with ten different random seeds from 0 to 9. We show the result with $X$=300 with a random seed of 3 in the paper, and other visualized results are put in Appendix of replication package~\cite{anonymousRepo} due to space limitation. The following findings and conclusions hold for different $X$ and random seeds. Second, we feed sampled pairs including code snippets and queries to well-trained \Our{} and \unixcoder individually and obtain their representations. Third, we apply T-SNE~\cite{van2008visualizing} to reduce the dimensionality of obtained representations into 2D and visualize dimensionality-reduced representations in~\Fig~\ref{fig:t-sne}.  In detail, \Fig~\ref{fig:t-sne-unixcoder} and \Fig~\ref{fig:t-sne-cocosoda} show representations learned by \unixcoder{} and \Our{}, respectively. The code snippet and query are marked in orange and blue, respectively. The distance between the paired code and query is indicated by a green line. From the~\Fig~\ref{fig:t-sne}, we can see that \textit{(1)} there are many very short green lines in \Fig~\ref{fig:t-sne} (a) and (b), which means that both \Our{} and \unixcoder can map most paired code snippets and queries into close embeddings. \textit{(2)} \Fig~\ref{fig:t-sne-unixcoder} have more long green lines than \Fig~\ref{fig:t-sne-cocosoda}. It indicates that \unixcoder has more cases than \Our{}, where pairwise representations of queries and code snippets are far away from each other. In the future, we will conduct error analysis to study the paired samples with long green lines. In summary, the visualization results intuitively show that \Our{} can learn better representations than \unixcoder.

\begin{tcolorbox}
\textbf{Summary.} Our model can learn a more uniform distribution of the representations of unimodal data (code snippets or queries), and the learned representations of paired code snippets and queries can be well aligned. 
\end{tcolorbox}

\section{Discussion}
\label{discussion}

\subsection{The Performance of \Our{} Without Being Fine-Tuned}
\label{sec:RQ_not-fine}
\begin{table}[t]
 \centering
\tablesize
\setlength{\tabcolsep}{2pt}
\caption{The performance of different approaches under the zero-short experimental setting evaluated by MRR scores.}
\begin{tabular}{lccccccc} 
\toprule
Model & \ruby{} & JS &\go  & \python & \java & \php & \avg \\
\midrule
\codebert &0.002 &0.001 &0.002 &0.001 &0.001 &0.001 &0.001 \\
\codetf &0.006 &0.003 &0.009 &0.001 &0.001 &0.001 &0.004 \\
\graphcodebert &0.238 &0.111 &0.209 &0.137 &0.123 &0.120 &0.156 \\
\unixcoder &0.576 &0.442 &0.648 &0.447 &0.466 &0.373 &0.492 \\
\midrule
\multirow{2}{*}{\Our{}} &\textbf{0.786} &\textbf{0.709} &\textbf{0.881} &\textbf{0.696} &\textbf{0.717} &\textbf{0.640} &\textbf{0.738} \\
&$\uparrow$36.5\% &$\uparrow$60.4\% &$\uparrow$36.0\% &$\uparrow$55.7\% &$\uparrow$53.9\% &$\uparrow$71.6\% &$\uparrow$50.0\% \\
\bottomrule
\end{tabular}
\label{tab:zero_short}
\end{table}

To further study the effectiveness of \Our{}, we evaluate different pre-trained models under the zero-short experimental setting, where models are directly used for evaluation without fine-tuning them. The experimental results are shown in \Tab~\ref{tab:zero_short}. We can see that pre-trained \codebert, \codetf, and \graphcodebert without fine-tuning performs poorly due to the representation degeneration problem~\cite{LiZHWYL20,GaoHTQWL19}. That is, 
the high-frequent tokens dominate the sequence representation~\cite{LiZHWYL20}, resulting in the poor sequence-level semantic representation of the code snippet and query. \unixcoder and our models adopt the contrastive learning related technique to obtain better representations of code snippets and queries, and perform better than the other two models. Especially, our approach can help the model to learn a uniform distribution of representation of unimodal data, and learn a good alignment for multimodal data. Therefore, our model outperforms other pre-trained models by about 50\% on average MRR scores even more than 70\% on \php language. Furthermore, the performance of \Our{} under zero-short setting even exceeds many fine-tuned models such as \codebert, \codetf and \graphcodebert (\Tab~\ref{tab:comp_with_baselines}) on average MRR. In summary, 
our pre-trained model performs better than other pre-trained models when all of them are not fine-tuned. Furthermore, the performance of \Our{} without being fine-tuned is even better than many fine-tuned models on average MRR.

\subsection{Limitations \& Threats to Validity}
Although \Our{} has an overall advantage, our model could still return inaccurate results, especially for the code snippets that use the third-library API or self-defined methods. This is because \Our{} only considers the information of the code snippet itself rather than other contexts such as other methods in the enclosing class or project~\cite{wang2020cocogum, BansalHM21}.
In our future work, more contextual information (such as enclosing class/project and called API/methods) could be considered in our model to further improve the performance of \Our{}.

We also identify the following threats to our approach:

\emph{Programming Languages.} 
Due to the heavy effort to evaluate the model on all programming languages, we conduct our experiment with as many programming languages as possible on the existing build datasets. Our model on different programming languages would have different results. In the future, we will evaluate the effectiveness of our approach with more other programming languages.

\emph{Pre-trained models.} To demonstrate that our approach is orthogonal to the pre-trained technique on the performance improvements for code search, we have adopted and evaluated our approach on four pre-trained models including a natural language pre-trained model \roberta and three source code pre-trained models  \codebert,  \graphcodebert and \unixcoder. It remains to be verified whether or not the proposed approach is applicable to other pre-trained models such as
GPT~\cite{BrownMRSKDNSSAA20} and T5~\cite{0034WJH21}.

\emph{Evaluated benchmark.}
The paired code snippet is usually used as the correct result for the given query. In fact, some unpaired code snippets also answer the given query. In the future, we will invite some developers to manually score the semantical correlation between the arbitrary code snippet and query and build a high-quality code search benchmark.

\section{Conclusion}
\label{conclusion}
 In this paper, we present \Our{}, which leverages multimodal momentum contrastive learning and soft data augmentation for code search. It can help the model learn effective representations by pulling together representations of code-query pairs and pushing apart the unpaired code snippets and queries. We conduct extensive experiments on a large-scale benchmark dataset with six programming languages and the results confirm its superiority. In our future work, more contextual information (such as enclosing class/project) could be considered in our model to further improve the performance of \Our{}. Replication package including datasets,  source code, and Appendix is available at~\revise{\url{https://github.com/DeepSoftwareAnalytics/CoCoSoDa}}.

\section*{Acknowledgement}
We thank reviewers for their valuable comments on this work. This research was supported by National Key R\&D Program of China (No. 2017YFA0700800) and Fundamental Research Funds for the Central Universities under Grant xtr072022001.
We would like to thank Jiaqi Guo for their valuable suggestions and feedback during the work discussion process. 

\bibliographystyle{IEEEtran}
\bibliography{ref}

\end{document}

%% file: tool.tex
\usepackage{enumitem}
\usepackage{makecell}
\usepackage{xspace}
\usepackage{nicematrix}
\usepackage{amsmath,amsfonts}

\usepackage{amsmath,amssymb,amsfonts}
\usepackage{graphicx}
\usepackage{textcomp}
\usepackage{subfigure}
\usepackage{tcolorbox}
\usepackage{booktabs}
\usepackage{tabularx}
\usepackage{lipsum}
\usepackage{multirow}

\newcommand{\myauthornote}[3]{}

\newcommand{\later}[1]{}
\newcommand{\revise}[1]{{\color{black} #1}}

\newcommand{\java}{Java\xspace}
\newcommand{\ruby}{Ruby\xspace}
\newcommand{\go}{Go\xspace}

\newcommand{\php}{PHP\xspace}
\newcommand{\python}{Python\xspace}
\newcommand{\javascript}{JavaScript\xspace}
\newcommand{\avg}{Avg. \xspace}

\newcommand{\Our}{CoCoSoDa\xspace} 
\newcommand{\soda}{SoDa\xspace} 
\newcommand{\Masking}{dynamic masking\xspace}
\newcommand{\Replacement}{dynamic replacement\xspace} 
\newcommand{\TypeMasking}{dynamic masking of specified type\xspace} \newcommand{\TypeReplacement}{dynamic replacement of specified type\xspace}
\newcommand{\ShortMasking}{DM\xspace}
\newcommand{\ShortReplacement}{DR\xspace} 
\newcommand{\ShortTypeMasking}{DMST\xspace} \newcommand{\ShortTypeReplacement}{DRST\xspace}

\newcommand{\nbow}{NBow\xspace}
\newcommand{\cnn}{CNN\xspace}
\newcommand{\birnn}{BiRNN\xspace}

\newcommand{\selfann}{SelfAtt\xspace}
\newcommand{\roberta}{RoBERTa\xspace}
\newcommand{\robertaOur}{\Our\textsubscript{RoBERTa}\xspace}
\newcommand{\codebert}{CodeBERT\xspace} 
\newcommand{\unixcoder}{UniXcoder\xspace} 
\newcommand{\codetf}{CodeT5\xspace} 
\newcommand{\plbart}{PLBART\xspace} 
\newcommand{\sptcode}{SPT-Code\xspace} 
\newcommand{\syncobert}{SyncoBERT\xspace} 
\newcommand{\codebertOur}{\Our\textsubscript{CodeBERT}\xspace} 
\newcommand{\graphcodebert}{GraphCodeBERT\xspace}
\newcommand{\graphcodebertOur}{\Our\textsubscript{GraphCodeBERT}\xspace}
\newcommand{\graphcodebertOurShort}{\Our\textsubscript{Graph}\xspace}
\newcommand{\sigircon}{Corder\xspace}

\newcommand{\Fig}{Fig.\xspace}

\newcommand{\Tab}{Table\xspace}
\newcommand{\Sec}{Sec.\xspace}
\newcommand{\Eq}{Eq.\xspace}

\usepackage{CJKutf8}
\usepackage[utf8]{inputenc} 

\usepackage{listings}
\usepackage{xcolor}
\definecolor{light-gray}{gray}{0.99}
\usepackage{fancybox}


\usepackage{arydshln}

\newcommand{\tablesize}{\scriptsize}